\documentclass[ALICE,manyauthors]{cernphprep}
\usepackage{AliHFMEXeXe}

\DeclareUnicodeCharacter{00A0}{}

\begin{document}

\begin{titlepage}
\PHyear{2020}
\PHnumber{208}        
\PHdate{06 November}  

\title{Inclusive heavy-flavour production at central and forward rapidity in Xe--Xe collisions at $\sqrt{s_{\rm NN}}=5.44$~TeV}
\ShortTitle{Inclusive heavy-flavour production in Xe--Xe collisions at $\sqrt{s_{\rm NN}}=5.44$~TeV}  

\Collaboration{ALICE Collaboration\thanks{See Appendix~\ref{app:collab} for the list of collaboration members}}
\ShortAuthor{ALICE Collaboration} 

\begin{abstract}

The first measurements of the production of muons and electrons from heavy-flavour hadron decays in Xe--Xe collisions at $\sqrt{s_{\rm NN}}$ = 5.44 TeV, using the ALICE detector at the LHC, are reported. The measurement of the nuclear modification factor $R_{\rm AA}$ 
is performed as a function of transverse momentum $\pT$ in several centrality classes at forward rapidity ($2.5 < y <4$) and midrapidity ($\vert y \vert < 0.8$) for muons and electrons from heavy-flavour hadron decays, respectively. A suppression by a factor up to about 2.5 compared to the binary-scaled pp reference is observed in central collisions at both central and forward rapidities. The $R_{\rm AA}$ of muons from heavy-flavour hadron decays is compared to previous measurements in Pb--Pb collisions at $\sqrt{s_{\rm NN}}$ = 5.02 TeV. 
When the nuclear modification factors are compared in the centrality classes 0--10\% for Xe--Xe collisions and 10--20\% for Pb--Pb collisions, which have similar charged-particle multiplicity density, a similar suppression, with $\RAA \sim 0.4$ in the $\pT$ interval $4 < p_{\rm T} < 8$ GeV/$c$, is observed. The comparison of the measured $R_{\rm AA}$ values in the two collision systems brings new insights on the properties of the quark-gluon plasma by investigating the system-size and geometry dependence of medium-induced parton energy loss. The results of muons and electrons from heavy-flavour hadron decays provide new constraints to model calculations.

\end{abstract}
\end{titlepage}

\setcounter{page}{2}


\section{Introduction}\label{sec1}

A state of strongly-interacting matter in which quarks and gluons are deconfined, the quark--gluon plasma (QGP), is created in ultra-relativistic heavy-ion collisions at high energy density and high temperature~\cite{Muller:2006ee,Bazavov:2011nk,pQCD2,Bazavov:2014pvz,Borsanyi:2013bia}. Due to their large masses ($m_\text{c}\approx 1.5~\mbox{GeV/}c^2$, $m_\text{b}\approx 4.8 \mbox{ GeV/}c^2$~\cite{Zyla:2020zbs}), 
charm and beauty quarks (heavy quarks) are mostly produced via  hard partonic scattering processes in the initial stages of the collision, with typical time scales smaller than the QGP formation time ($\sim$ 1 fm/$c$~\cite{Liu:2012ax}). Furthermore, additional thermal production, as well as annihilation rates, of charm and beauty quarks in the strongly-interacting matter are expected to be small even at LHC energies \cite{bib::Averbeck,bib::Braun_Munziger}.
Therefore, charm and beauty quarks are a valuable probe for a detailed characterisation of this deconfined medium.
During their propagation through the QGP, they lose energy via radiative~\cite{intro1,Radiativeb} and collisional~\cite{Colla,Collc} processes due to the microscopic interactions with the medium.

In order to quantify the effect of parton energy loss, we employ the nuclear modification factor $\RAA$ defined as the ratio of the $\pT$- and y-differential particle yield ($\mathrm{d}^2N_{\mathrm{AA}}/\mathrm{d}p_{\mathrm{T}}\mathrm{d}y$) in nucleus-nucleus collisions of a given centrality to the corresponding $\pT$- and $y$-differential production cross section ($\mathrm{d}^2\sigma_{\mathrm{pp}}/\mathrm{d}p_{\mathrm{T}}\mathrm{d}y$) in pp collisions at the same centre-of-mass energy, scaled by the average nuclear overlap function $\langle T_{\rm AA} \rangle$:

\begin{equation}
    R_{\mathrm{AA}}(p_\mathrm{T},y) = 
    \frac{1}{\langle T_{\mathrm{AA}}\rangle}\cdot\frac{\mathrm{d}^2N_{\mathrm{AA}}/\mathrm{d}p_{\mathrm{T}}\mathrm{d}y}{\mathrm{d}^2\sigma_{\mathrm{pp}}/\mathrm{d}p_{\mathrm{T}}\mathrm{d}y}\hspace{1mm},
    \label{eq:RAA}
\end{equation}

where $\langle T_{\text{AA}}\rangle$ is defined, for a given centrality class, as the average number of binary nucleon-nucleon collisions per A--A collision in that class, divided by the inelastic nucleon-nucleon cross section~\cite{Glauber:1970jm, Miller:2007ri,ALICE-PUBLIC-2018-011}.

A wealth of results on the measurement of open heavy-flavour production at the LHC was obtained in Pb--Pb collisions at $\sNN$ = 2.76 TeV and 5.02 TeV. The results from the ALICE, ATLAS and CMS Collaborations show a clear evidence of a strong suppression of open heavy-flavour yields compared to the binary-scaled pp reference in central collisions (see~\cite{Andronic:2015wma} and references therein, and~\cite{Adam:2015nna, Sirunyan:2017xss, Acharya:2018hre, Sirunyan:2017oug, Aaboud:2018bdg,Acharya:2019mom,pubPbPb} for recent publications). The measurements of the production of electrons and muons from heavy-flavour hadron decays at midrapidity ($|y| <$ 0.8) and forward rapidity ($2.5 < y < 4$), respectively, show a suppression up to a factor of about 3 in the 10\% most central collisions~\cite{Acharya:2019mom,pubPbPb}. The nuclear modification factor of electrons from charm and beauty-hadron decays \cite{Adam:2016wyz, Adam:2015qda} and of \textit{D} mesons \cite{Abelev:2014hha} in p--Pb collisions at
$\sqrt{s_\mathrm{NN}}=5.02\text{ TeV}$  was found to be consistent with unity within uncertainties.
From this, one can conclude that the strong suppression observed in Pb--Pb collisions is due to substantial final-state interactions of heavy quarks with the QGP formed in heavy-ion collisions.

The study of open heavy-flavour production in different collision systems with similar collision energies can provide more insight on the role of initial state effects, like the modification of the parton distribution functions (PDF) inside bound nucleons \cite{Eskola:2009uj}, and on the production mechanisms of final-state particles, like hadronisation via fragmentation and coalescence \cite{Greco:2003vf, Andronic:2015wma}. Furthermore, such measurements can be used to test the path-length dependence of the heavy-quark energy loss \cite{Djordjevic:2018ita}. Previous measurements of the system-size dependence of open heavy-flavour production were performed at RHIC, comparing results from Au--Au and Cu--Cu collisions, both at $\sNN$
~=~200 GeV. The PHENIX collaboration measured a significant suppression of muons from heavy-flavour hadron decays at forward rapidity (1.4 $< y <$ 1.9) for $\pT > 2$~GeV/$c$ in the 20\% most central Cu--Cu collisions at $\sNN$ = 200 GeV~\cite{PHENIXRaamuCuCu}. The nuclear modification factor of electrons from heavy-flavour hadron decays measured at midrapidity ($\vert  y \vert < 0.35$) in Cu--Cu collisions at $\sNN$ = 200 GeV/$c$ is found to be in agreement with that obtained in Au--Au collisions at $\sNN$ = 200 GeV/$c$, when the two are compared at similar number of participating nucleons $\langle N_{\rm part} \rangle$~\cite{Adare:2013yxp}. The LHC delivered for the first time collisions of xenon ions (${\rm ^{129}Xe^{54+}}$) at a centre-of-mass energy per nucleon pair $\sNN$ = 5.44 TeV during a pilot run of 6 hours at the end of 2017.  This allows one to complement the findings in Pb--Pb collisions and to study the dependence of particle production on the size of the collision system and of the produced medium. A remarkably similar suppression was observed in Pb--Pb and Xe--Xe collisions for charged particles, in terms of their nuclear modification factor compared in event classes with similar charged-particle multiplicity density $\langle$d$N_{\rm ch}$/d$\eta\rangle$, possibly indicating similar medium densities and sizes in the two collision systems under that condition~\cite{Sirunyan:2018eqi, Acharya:2018eaq}. The CMS collaboration reported a slightly smaller $\RAA$ for charged particles measured in Xe--Xe collisions than in Pb--Pb collisions when the comparison is performed for centrality classes with a similar number of participating nucleons~\cite{Sirunyan:2018eqi}. A good agreement was also found between the $\RAA$ of inclusive $\rm J/\psi$  measured in Pb--Pb and Xe--Xe within uncertainties~\cite{Acharya:2018jvc}. 

This letter presents the first measurement of the production of open heavy-flavour hadrons via the muon and electron decay channels at forward rapidity ($2.5 < y < 4$) and midrapidity ($\vert y \vert < 0.8$) in Xe--Xe collisions at $\sNN$ = 5.44 TeV with the ALICE detector at the LHC. The $\pT$-differential yield and the nuclear modification factor are presented in several centrality intervals expressed in terms of percentages of the inelastic Xe--Xe cross section, namely 0--10\%, 10--20\%, 20--40\% and 40--60\% (0--20\% and 20--40\%) for the muon (electron) analysis. The upper limit of the $\pT$ range available for the measurements sits between 6 and 8 GeV/$c$, depending on the centrality class and decay channel. 
In the midrapidity region, electrons from heavy-flavour hadron decays are measured, for the first time at the LHC, down to $\pT$ = 0.2~GeV/$c$ in the 20--40\% centrality interval thanks to the reduced magnetic field of 0.2~T in the ALICE solenoid magnet, as compared to the nominal field of 0.5~T for Pb--Pb collisions. 
The comparison between Pb--Pb collisions and the smaller Xe--Xe system is extended to the open heavy-flavour sector with the measurement of the $\RAA$ of muons from heavy-flavour hadron decays at forward rapidity in both systems. Comparisons with transport model predictions are reported. The measurements can add to our understanding of the initial state and the parton in-medium energy loss mechanisms. In particular, the results discussed in this letter can bring additional constraints on the model parameters sensitive to the path-length dependence of in-medium energy loss

\section{Experimental apparatus}\label{sec2} 

The ALICE detector is described in detail in \cite{Aamodt:2008zz,bib::Alice_performance} and references therein. The apparatus consists of a central barrel at midrapidity with pseudorapidity coverage $|\eta|<0.9$, a muon spectrometer at forward rapidity ($-4<\eta<-2.5$) and a set of detectors for triggering and event characterisation installed in the forward and backward rapidity regions. The central barrel detectors are embedded in a solenoid, which provided a magnetic field of 0.2 T parallel to the beam direction during the Xe--Xe data taking.

At midrapidity, the Inner Tracking System (ITS) and the Time Projection Chamber (TPC) are used for track reconstruction. The ITS detector consists of six cylindrical silicon layers surrounding the beam pipe: the Silicon Pixel Detector (SPD), the Silicon Drift Detector (SDD) and the Silicon Strip Detector (SSD). Along with the momentum measurement, the TPC, the SDD and the SSD also provide specific ionisation energy loss (d$E$/d$x$) information, used for charged-particle identification (PID). The particle identification is completed by the information provided by a Time-Of-Flight (TOF) detector based on multi-gap resistive plate chambers.

The muon spectrometer~\cite{CERN-LHCC-2000-046}, as seen from the interaction point, consists of a 10 nuclear interaction length ($\lambda_{\rm I}$) front absorber filtering hadrons, electrons, photons and reducing the yield of muons from light-flavour hadron decays, five tracking stations with the third one placed in a dipole magnet with a field integral of 3 T$\cdot$m, a 1.2~m thick iron wall (7~$\lambda_{\rm I}$) and two trigger stations. Each tracking station is composed of two planes of cathode pad chambers. The iron wall stops secondary hadrons escaping from the front absorber as well as residual low momentum muons from light-hadron decays. Each trigger station consists of two planes of resistive plate chambers. A conical absorber protects the muon spectrometer throughout its full length against secondary particles produced by the interaction with the beam pipe of primary particles at large $\eta$ values.

The V0 detector consists of two arrays of 32 scintillator tiles each, covering the pseudorapidity intervals $2.8<\eta<5.1$ (V0A) and $-3.7<\eta<-1.7$ (V0C). It serves as trigger detector and is used for the centrality estimation. The latter is performed through a Glauber Monte Carlo (MC) fit of the signal amplitude in the two scintillator arrays~\cite{Abelev:2013qoq,ALICEDRPbPbcent}. The centrality intervals are defined as percentiles of the hadronic Xe--Xe 
cross section~\cite{Loizides:2017ack}. The Zero Degree Calorimeters (ZDC), located on both sides of the interaction point at $z = \pm$~112.5 m, are used, both online and offline, for the rejection of electromagnetic interactions and the timing information from both the V0 and ZDC is employed offline to reject beam-induced background. 

The Forward Multiplicity Detector (FMD), made of three sets of silicon strip sensors covering the $\eta$ intervals $-3.5 < \eta < -1.8$ and $1.8 < \eta <5$, provides a measurement of the charged-particle multiplicity density at forward rapidity. 

The minimum bias (MB) trigger required at least a hit in each of the V0A and V0C, and at least one neutron detected by the ZDC on each side of the interaction point. The measurement of the production of electrons from heavy-flavour hadron decays at midrapidity was performed on a sample of events characterized by the MB trigger condition. In addition, only events with a reconstructed primary vertex with $|v_{\rm z}|~<$~10 cm, where $v_{\rm z}$ is the longitudinal coordinate along the beam axis, are used in the analysis. 

A sample of muon-triggered events was considered for the measurement of muons from heavy-flavour hadron decays. The muon trigger condition~\cite{Aamodt:2008zz} required 
the coincidence of at least one hit in each of the V0A and V0C, and at least one track segment in the muon trigger system, with a $\pT$ above the threshold of the online trigger algorithm. 
The trigger setting provides a  $\sim$ 50\% efficiency for muons with $\pT$~$\approx$~0.5~GeV/$c$. 

The probability for multiple interactions in a single bunch crossing (pile-up) was negligible (about $10^{-4}$) for both event samples, due to the very low interaction rate. The number of events analysed with different trigger samples and centrality intervals is summarised in Table~\ref{tab::num_evs}, together with the corresponding values of the average nuclear overlap function $\langle T_\mathrm{AA}\rangle$ \cite{ALICE-PUBLIC-2018-011,Loizides:2017ack}. The integrated luminosity amounts to about 0.30~$\mu\rm b^{-1}$ and 0.33~$\mu\rm b^{-1}$ for the MB- and muon-triggered data samples.

\begin{table}[ht!]
	\centering
	\caption{
	Number of analysed Xe--Xe events in different centrality intervals and trigger configurations. The corresponding $\langle T_\mathrm{AA}\rangle$ values are also reported. The MB and muon-triggered samples are used for the measurement of electrons from heavy-flavour hadron decays at midrapidity and muons from heavy-flavour hadron decays at forward rapidity.
	}
	\begin{tabular}{lccc}
        \hline 
			 centrality & MB trigger & muon trigger & $\langle T_\mathrm{AA}\rangle$ (mb$^{-1}$)\\
			  & (electrons) & (muons) & \\
		\hline 
0--10\% & -- & $1.79 \cdot 10^{5}$ & $12.33 \pm 0.71$\\
 10--20\% &-- &  $1.62 \cdot 10^{5}$ & $7.465 \pm 0.520$ \\
  0--20\%  & $0.3 \cdot 10^{6}$ & -- & $9.896 \pm 0.620$ \\
		 20--40\%& 0.3 $\cdot 10^{6}$ &  $2.48 \cdot 10^{5}$ & $3.466 \pm 0.350$\\
40--60\%  & -- &  $1.29 \cdot 10^{5}$ & $1.008 \pm 0.131$\\
    \hline 
	\end{tabular}
	\label{tab::num_evs}
\end{table}

\section{Data analysis} 
\label{sec3}

The $\pT$-differential yield of electrons (muons) from semi-leptonic heavy-flavour hadron decays is obtained by measuring the inclusive yields and subtracting the contribution of electrons (muons) that do not originate from open heavy-flavour hadron decays. 

\subsection{Muons from heavy-flavour hadron decays}
\label{sec:mudata}

Standard criteria are applied to select track candidates reconstructed with the muon spectrometer using the tracking algorithm described in~\cite{Aamodt:2011gj}. Tracks are required to be within the pseudorapidity interval $-4 < \eta < -2.5$ and their polar angle at the end of the front absorber ($\theta_{\rm abs}$) should satisfy the condition $170^\circ < \theta_{\rm abs} < 178^\circ$. The $\theta_{\rm abs}$ selection reduces the impact of multiple scattering by rejecting tracks crossing the high-density region of the front absorber. Reconstructed tracks that match a track segment in the muon trigger system are identified as muons. 
Finally, the  distance of the track to the primary vertex (DCA) weighted by its momentum ($p$) is required to be smaller than 6 times the width of the weighted DCA distribution obtained considering all reconstructed tracks~\cite{Alice:2016wka}. This selection reduces the fraction of fake tracks and beam-induced background tracks. 
After the selection criteria are applied, the main background contributions in the region $3 < \pT < 8$~GeV/$c$ which need to be estimated and further subtracted consist of i) muons from primary charged-pion and charged-kaon decays, ii) muons from secondary charged-pion and charged-kaon decays originating from the interaction of light-charged hadrons with the material of the front absorber and iii) muons from $\rm J/\psi$ decays. Therefore, the differential yields of muons from heavy-flavour hadron decays in a given centrality class are obtained as
\begin{linenomath}
\begin{equation}
{{{\rm d^2}N^{\mu^\pm\leftarrow {\rm {HF}}}} \over
{{\rm d}p_{\rm T} {\rm d}y }} = 
{{{\rm d}^2N^{\rm \mu^\pm}}\over {{\rm d}p_{\rm T}{\rm d}{y}}}
- {{{\rm d}^2N^{\mu^\pm\leftarrow {\rm \pi}}}\over {{\rm d}p_{\rm T}{\rm d}{y}}}
- {{{\rm d}^2N^{\rm \mu^\pm\leftarrow {\rm K}}}\over {{\rm d}p_{\rm T}{\rm d}{y}}}
- {{{\rm d}^2N^{\rm \mu^\pm\leftarrow {{\rm sec.} {\rm \pi/K}}}}\over {{\rm d}p_{\rm T}{\rm d}y}}
- {{{\rm d}^2N^{\rm \mu^\pm\leftarrow {\rm J/\psi}}}\over {{\rm d}p_{\rm T}{\rm d}{y}}},
\label{eq:CrossSecHF}
\end{equation}
\end{linenomath}
where ${\rm d^2}N^{\mu^\pm}/{\rm d}p_{\rm T} {\rm d}y$ stands for the differential yield of inclusive muons corrected for acceptance, and tracking and trigger efficiency ($A \times \epsilon$), and normalised to the equivalent number of MB events, while ${\rm d}^2N^{\mu^\pm\leftarrow {\rm \pi}}/{\rm d}p_{\rm T}{\rm d}{y}$, ${\rm d}^2N^{\rm \mu^\pm\leftarrow {\rm K}}/{\rm d}p_{\rm T}{\rm d}{y}$, ${\rm d}^2N^{\rm \mu^\pm\leftarrow {{\rm sec.} {\rm \pi/K}}}/{\rm d}p_{\rm T}{\rm d}y$ and ${\rm d}^2N^{\rm \mu^\pm\leftarrow {\rm J/\psi}}/{\rm d}p_{\rm T}{\rm d}{y}$ represent the corresponding estimated differential yields of muons from charged-pion decays, muons from charged-kaon decays, muons from secondary charged-pion and kaon decays and muons from $\rm J/\psi$ decays. 

In order to evaluate the equivalent number of MB events in the triggered-muon sample a normalisation factor $F_{\rm norm}$ needs to be determined. This factor corresponds to the inverse probability of having a triggered muon in a MB event. It is calculated using two different procedures as detailed in~\cite{Acharya:2019mky,pubPbPb}, either applying the muon trigger condition in the analysis of MB events or from the relative trigger rates. The obtained value is $F_{\rm norm} = 2.428 \pm 0.024$. The quoted uncertainty is systematic, the statistical one is found to be negligible. The systematic uncertainty is obtained by comparing the results obtained from the two methods.  

The correction factors for $A \times \epsilon$ are estimated in the same way as in~\cite{pubPbPb}, using simulated muons from heavy-flavour hadron decays generated with $\pT$ and $y$ distributions based on a fixed order with next-to-leading-log resummation model (FONLL) \cite{Cacciari:1998it} and a realistic description of the detector conditions. The effect of the detector occupancy on the muon tracking efficiency as a function of centrality in Pb--Pb collisions was studied in~\cite{pubPbPb} by embedding simulated muons from heavy-flavour hadron decays in real MB-triggered events. This analysis uses the centrality-dependent efficiency scaling factors obtained from simulation in~\cite{pubPbPb} for Pb--Pb centrality intervals with similar charged-particle multiplicity density at forward rapidity ($\langle {\rm d}N_{\rm ch}/{\rm d}\eta \rangle_{2.5 < y < 4}$)~\cite{Adam:2015ptt,Acharya:2018hhy} as the considered Xe--Xe centrality intervals. This results in a $\sim$3\% decrease of the efficiency in the 10\% most central Xe--Xe collisions compared to the most peripheral collisions (60--80\% centrality interval), independent of $\pT$. The obtained values of $A \times \epsilon$ vary between between 84\% and 87\%, depending on the selected centrality class, with no significant $\pT$ dependence in the interval of interest.

The determination of the contribution of muons from primary charged-pion and charged-kaon decays relies on a data-driven simulation of the different contributions, as described later. Since the charged-pion and charged-kaon spectra are not available, the following strategy is developed. The procedure is based on the $\pi^\pm$ and K$^\pm$ spectra measured with ALICE at midrapidity up to $\pT$~=~20 GeV/$c$ in 
Pb--Pb collisions at $\sNN$~=~5.02 TeV for various centrality intervals~\cite{Acharya:2019yoi}, which are scaled by the measured ratio of the charged-particle $\pT$ distribution in Xe--Xe collisions to that in Pb--Pb collisions~\cite{Acharya:2018eaq,Acharya:2018qsh}. 
The obtained spectra in Xe--Xe collisions at $\sNN$~=~5.44~TeV are used as input for the background estimation. The extrapolation to the forward rapidity region assumes that the $\RAA$ of charged pions and charged kaons is independent of rapidity up to large $y$, $y > 4$ \cite{pubPbPb}. 
This leads to
\begin{linenomath}
\begin{equation}
\bigg\lbrack \frac {{\rm d}^2 N^{\pi^\pm (K^\pm)}} 
{{\rm d}p_{\rm T}{\rm d} y} \bigg\rbrack_{\rm AA} = 
\langle N_{\rm coll}\rangle \cdot 
\lbrack R_{\rm AA}^{\pi^\pm (K^\pm)} \rbrack^{\rm mid-y}\cdot 
\lbrack F^{\rm \pi^\pm (K^\pm)}_{\rm extrap} (p_{\rm T}, y) \rbrack_{\rm pp} \cdot
 \bigg\lbrack \frac {
{\rm d}^2 N^{\pi^\pm (K^\pm)}} {{\rm d}p_{\rm T} {\rm d} y}
\bigg\rbrack ^{{\rm mid}-y}_{\rm pp}.
\label{eq:RAAextrap1}
\end{equation}
\end{linenomath}
Finally, replacing the midrapidity nuclear modification factor $\lbrack R_{\rm AA}^{\pi^\pm (K^\pm)} \rbrack^{\rm mid-y}$ by its expression, Eq.~(\ref{eq:RAAextrap1}) simplifies as
\begin{linenomath}
\begin{equation}
\bigg\lbrack \frac {{\rm d}^2 N^{\pi^\pm (K^\pm)}} 
{{\rm d}p_{\rm T} {\rm d} y} \bigg\rbrack_{\rm AA} = 
\lbrack F^{\rm \rm \pi^\pm (K^\pm)}_{\rm extrap} (p_{\rm T}, y) \rbrack_{\rm pp} \cdot
 \bigg\lbrack \frac {
{\rm d}^2 N^{\pi^\pm (K^\pm)}} {{\rm d}p_{\rm T} {\rm d} y}
\bigg\rbrack ^{{\rm mid}-y}_{\rm AA}.
\label{eq:RAAextrap2}
\end{equation}
\end{linenomath}

In these equations,  $\bigg\lbrack \frac {{\rm d}^2 N^{\pi^\pm (K^\pm)}} {{\rm d}p_{\rm T} {\rm d} y}\bigg\rbrack ^{{\rm mid}-y}_{\rm AA}$ and $\bigg\lbrack \frac {{\rm d}^2 N^{\pi^\pm (K^\pm)}} {{\rm d}p_{\rm T} {\rm d} y} \bigg\rbrack_{\rm AA}$ are the midrapidity spectra in Xe--Xe collisions and the $y$-extrapolated ones, respectively. The $\pT$-dependent rapidity factor $\lbrack F^{\rm \rm \pi^\pm (K^\pm)}_{\rm extrap} (p_{\rm T}, y) \rbrack_{\rm pp}$, used to extrapolate the midrapidity $\pi^\pm$ and $\rm K^\pm$ spectra in pp collisions to the forward rapidity region, is estimated by means of pp simulations at $\sqrt{s}$ = 5.44 TeV using PYTHIA 8.2~\cite{Sjostrand:2014zea}\footnote{
It was checked that simulations with PYTHIA 6.4 and PHOJET, which were also performed for the rapidity extrapolation~\cite{Acharya:2019mky}, give compatible results
within uncertainties for the rapidity extrapolation.}. The procedure employed to perform the rapidity extrapolation and to account for its $\pT$ dependence is described in~\cite{Acharya:2019mky,pubPbPb}. The systematic uncertainty due to the assumption that the $\RAA$ remains unchanged from midrapidity up to $y = 4$ will be presented in section \ref{syst}. 
Finally, the $\pT$ and $y$ distributions of decay muons are generated via a fast detector simulation of the decay kinematics and absorber effects, using the 
$y$-extrapolated charged-pion and charged-kaon distributions. For each centrality class, the yields are normalized to the equivalent number of MB events and subtracted from the inclusive muon distribution. 
The estimated fraction of muons from charged-pion and charged-kaon decays depends on $\pT$ and collision centrality. It decreases from about 14\% at $\pT$ = 3~GeV/$c$ down to 5\% at $\pT$~=~8~GeV/$c$ in the 0--10\% centrality class, while in the 40--60\% centrality class it decreases from about 20\% to 8\% for $\pT$ increasing from 3 GeV/$c$ to 7~GeV/$c$.

The contribution of muons from secondary light-hadron decays is estimated by means of HIJING v1.383 simulations~\cite{Hijing:ref} with the GEANT3 transport code~\cite{Brun:1082634}. These simulations indicate that the fraction of muons from secondary $\pi^\pm$ and $\rm K^\pm$ decays with respect to muons from primary $\pi^\pm$ and $\rm K^\pm$ decays amounts to about 9\%, independently of $\pT$ in the kinematic region of interest, and of the centrality selection. The fraction of muons from secondary $\pi^\pm$ and $\rm K^\pm$ decays is obtained from the estimated contribution of muons from primary  $\pi^\pm$ and $\rm K^\pm$ decays. The resulting fraction of muons from secondary $\pi^\pm$ and $\rm K^\pm$ decays with respect to inclusive muons decreases from about 1.3\% at $\pT$ = 3 GeV/$c$ to about 0.4\% at $\pT$ = 8 GeV/$c$ in the 0--10\% centrality class.  

Differential measurements of the J/$\psi$ production at forward rapidity are not available in Xe--Xe collisions~\cite{Acharya:2018jvc}. 
The estimation of the background component of muons from J/$\psi$ decays relies on the measurements in Pb--Pb collisions~\cite{pubPbPb} in centrality intervals with a comparable average charged-particle multiplicity as in Xe--Xe collisions. This is achieved by extrapolating the $\rm J/\psi$ $\pT$- and y-differential spectra measured by ALICE in the dimuon channel at forward rapidity ($2.5 < y < 4$) and $\pT < 12$~GeV/$c$~\cite{Adam:2016rdg,Acharya:2019iur} to a larger kinematic range. A power-law function is used for the $\pT$-dependence and a Gaussian function for the rapidity dependence.
Then, the decay muon distributions are estimated with a fast detector simulation. 
The fraction of muons from J/$\psi$ decays in the Pb--Pb collision centrality class corresponding to an average charged-particle multiplicity similar to that in central (0--10\%) Xe--Xe collisions, for $2.5 < y  < 4$, lies in the interval 1--4\%, depending on $\pT$. The maximum of the distribution of muons from J/$\psi$ decays is located at intermediate $\pT$ ($4 < \pT < 6$~GeV/$c$). 

After the subtraction of the background sources, the corrected and normalized yields of muons from heavy-flavour hadron decays in each $\pT$ interval are further divided by the width of the $\pT$ interval and the rapidity coverage.

\subsection{Electrons from heavy-flavour hadron decays}

Candidate electron tracks are required to fulfill basic selection criteria similar to those reported in~\cite{Acharya:2019mom}. The rapidity interval used in the analysis is restricted to $|y|~<$ 0.8 to exclude the edges of the detector acceptance, where the systematic uncertainties related to particle identification increase. Only tracks that have hits on both SPD layers are accepted, in order to reduce the contamination of electrons from photon conversions in the detector material. 
The electron identification is mainly based on the measurement of the specific ionisation energy loss in the TPC (d$E$/d$x$), similarly to the procedure followed in ~\cite{Acharya:2018upq,Adam:2016khe}. The selection variable is defined as the standard deviation of d$E$/d$x$ from the parameterised Bethe-Bloch \cite{BetheBloch, PhysRevD.98.030001} expectation value for electrons, expressed in units of the d$E$/d$x$ resolution, $n_{\sigma\textrm{,e}}^\textrm{TPC}$ \cite{bib::Alice_performance}.
In addition, the TOF and the ITS detectors are used to separate 
electrons from kaons and protons, imposing the criteria $|n_{\sigma\textrm{,e}}^\textrm{TOF}|~<~3$ and $-4<n_{\sigma\textrm{,e}}^{\rm ITS}<2$ reduce the hadron contamination.
The $n_{\sigma\textrm{,e}}^\textrm{TOF}$ and $n_{\sigma\textrm{,e}}^{\rm ITS}$ are defined as the deviation of the time-of-flight in the TOF and the d$E$/d$x$ in the ITS from the expected values for electrons normalised to the respective detector resolutions.
The inclusive electron sample is finally extracted with the requirement $-1<n_{\sigma\textrm{,e}}^{\rm TPC}<3$, which corresponds to a 84\% efficiency. 
The residual hadron contamination is estimated fitting the full $n_{\sigma\textrm{,e}}^\textrm{TPC}$ distribution of the selected tracks with an analytic function including contributions for each particle species in different momentum intervals \cite{Acharya:2019hao}. The residual hadron contamination reaches a maximum value of 20\% ( 15\%) for momentum equal to 6 GeV/$c$ in the 0--20\% (20--40\%) centrality class. 

The selected inclusive electron sample does not only contain electrons from open heavy-flavour hadron decays, but also from different sources of background. The dominant background sources are electrons from Dalitz decays of light neutral mesons, mainly $\pi^0$ and $\eta$, and from photon conversions, which largely dominate the inclusive electron yield for $\pT~<$~1.5~GeV/$c$.
The ratio of the signal to the background electron contributions, measured in Xe--Xe collisions at $\sNN$~=~5.44~TeV, is about 0.2 at $\pT$~=~0.5~GeV/$c$ and is observed to increase with $\pT$, reaching a value of about 4 at $\pT~>$~3~GeV/$c$.
Other sources of background, like the di-electron decays of light vector mesons ($\rho_0,\omega,\phi$), electrons from $\mathrm{W}$ and $\mathrm{Z}/\gamma^{*}$ decays, di-electron decays of quarkonia and electrons from weak decays of kaons, are observed to be negligible in the $\pT$ interval of interest for this analysis \cite{Abelev:2012xe}.
The contribution of electrons from Dalitz decays of $\pi^0$ and $\eta$ and from photon conversions is estimated via an invariant mass analysis of unlike-sign pairs formed by pairing selected electron (positron) tracks with opposite-charge tracks identified as positrons (electrons). The latter are called associated electrons in the following. The combinatorial background is subtracted using the like-sign invariant mass distribution.
In order to maximise the probability to find the electron (positron) partner, 
the associated electrons are selected with looser criteria with respect to the ones applied for the inclusive electron selection~\cite{Acharya:2018upq,Adam:2016khe}.
Due to the detector acceptance and inefficiencies and the decay kinematics, not all Dalitz and conversion electrons in the inclusive electron sample are tagged with this method. Therefore, the raw yield of tagged electrons from the background sources is corrected for the efficiency to find the associated electron (positron), called tagging efficiency.
The generated $\pi^0$ $\pT$ distributions are weighted in order to match the measured $\pT$ spectra. The weights are defined as the product of the ratio of charged particles in Xe--Xe and Pb--Pb collisions and the charged-pion spectra measured in Pb--Pb collisions \cite{Acharya:2018eaq, Acharya:2018qsh, Acharya:2019yoi}. 
For the $\eta$ meson, the weights are determined via the $m_{\rm T}$-scaling \cite{Khandai:2011cf,Altenkamper:2017qot} of the computed pion-$\pT$ spectra.
The electron tagging efficiency increases with the electron $\pT$, starting from a value of about 40\% at $\pT$ = 0.2~GeV/$c$ and reaching a value of about 80\% at $\pT$ = 6~GeV/$c$.
It was observed in a previous analysis~\cite{Acharya:2018upq, Acharya:2019mom} that the contribution from J/$\psi$ decays reaches a maximum of about 4\% in the region $2<p_\mathrm{T}<3$ GeV/$c$ in central Pb--Pb collisions, decreasing to a few percent in more peripheral events. At lower and higher $\pT$, this contribution decreases quickly and becomes negligible. Hence, the contribution from J/$\psi$ decays is not subtracted in the present analysis and its contribution is accounted in the systematic uncertainties, as it was done for a previous measurement~\cite{Acharya:2019mom}.
Due to the significantly different level of suppression between heavy-flavour particles and kaons at low $\pT$ \cite{Adam:2015kca,Acharya:2019mom}, and due to the tight tracking requirements applied in the analysis, which suppress long-lived particle decays, the contribution from the weak decay ${\rm K^{0/\pm}} \rightarrow {\rm e}^{\pm}\pi^{\mp/0}\,\overset{\scriptscriptstyle(-)}{\nu_{e}}$ is expected to be negligible in this system, hence it is not subtracted.

After the statistical subtraction of the hadron contamination and background of electrons from Dalitz decays and from photon conversions, the obtained yields of electrons from heavy-flavour hadron decays in each $\pT$ interval are divided by i) the geometrical acceptance times the reconstruction and PID efficiencies, ii) the number of analysed events and iii) the width of the $\pT$ interval and of the covered $y$ interval in order to obtain the corrected and normalised differential yields of electrons from heavy-flavour hadron decays.
 
The efficiencies are computed using a Monte Carlo sample where the underlying Xe--Xe events are simulated using the HIJING v1.383 generator \cite{Hijing:ref} and the heavy-flavour signal, obtained using the PYTHIA6 generator~\cite{Sjostrand:2006za}, is embedded. The efficiency of
the TPC electron identification selection criterion is determined using a data-driven approach based on the $n_{\sigma\textrm{,e}}^\textrm{TPC}$ distribution \cite{Abelev:2012sca, Adam:2015qda}.
The total reconstruction efficiencies increase with $\pT$, starting from a value of about 10\% at $\pT$ = 0.2 GeV/$c$ and reaching a value of about 25\% at $\pT$~=~6~GeV/$c$.

\subsection{Systematic uncertainties} \label{syst}

In the muon analysis, the systematic uncertainties on the yield of muons from heavy-flavour hadron decays include contributions from the inclusive muon yield, the background contamination and the normalisation of muon-triggered events to the equivalent number of minimum bias events. 
The systematic uncertainty related to the inclusive muon yield is extensively discussed in~\cite{Acharya:2019mky,pubPbPb}. It is evaluated considering the following contributions: i) muon tracking efficiency (1\%), ii) muon trigger efficiency (2.2\%), including the intrinsic efficiency of the muon trigger chambers and the response of the trigger algorithm, iii) choice of the $\chi^2$ cut implemented in the matching between tracks reconstructed in the tracking chambers and trigger chambers (0.5\%) and iv) resolution and alignment of the tracking chambers (0.5\% for $ 7 < \pT < 8$~GeV/$c$ and negligible for $\pT < 7$~GeV/$c$). Finally, the systematic uncertainty arising from the procedure used to take into account the dependence of the muon tracking efficiency on the collision centrality, related to the detector occupancy, reaches 0.5\% in the 10\% most central collisions. 

A first contribution to the systematic uncertainty on the estimated contribution of muons from charged-pion (kaon) decays varying from about 5\% (9.5\%) at $\pT$ = 3 GeV/$c$ to 6\% (13\%) at $\pT$ = 8 GeV/$c$ in the 0--10\% centrality class comes from i) the measured midrapidity $\pT$ distributions of charged pions (kaons) in Pb--Pb collisions and ii) the ratio of the 
charged-particle spectrum measured at midrapidity in Xe--Xe collisions to that in Pb--Pb collisions. 
A second contribution of 9\% (6\%) for muons from charged-pion (kaon) decays is due to the rapidity extrapolation of the midrapidity pion and kaon distributions in pp collisions, needed to estimate the charged-pion (kaon) distributions in Xe--Xe collisions at forward rapidity. The $\pT$ dependence of the rapidity extrapolation introduces a maximum uncertainty of 3\% (1.5\%) at $\pT$ = 8 GeV/$c$ for muons from charged-pion (kaon) decays. These two systematic uncertainties are obtained by comparing the results from PYTHIA 8 with various color reconnection options~\cite{Christiansen:2015yqa, Acharya:2019mky}. A third contribution is attributed to the simulation of hadronic interactions in the front absorber and amounts to 4\%, independently of the mother particle. The total systematic uncertainty is obtained by adding in quadrature the contributions listed above, which gives a systematic uncertainty going from 10.5\% (12\%) at $\pT$ = 3 GeV/$c$ to 13\% (16\%) at $\pT$ = 8 GeV/$c$ for charged-pion (kaon) decay muons. 
Finally, the systematic uncertainty due to the assumption that the suppression pattern of $\pi^\pm$ and $\rm K^\pm$ is independent of rapidity up to $y = 4$ is estimated by changing the corresponding yields by $\pm 50\%$. The assigned uncertainty is the difference between the yields of muons from heavy-flavour hadron decays obtained in the two extreme cases, divided by $\sqrt{12}$, which corresponds to the RMS of a uniform distribution.
Moreover, in order to take into account possible effects of the transport code, the yields of muons from secondary charged-pion (kaon) decays are varied by $\pm$100\% and the difference between the yields of muons from heavy-flavour decays obtained in the two extreme cases, divided by $\sqrt{12}$, is assigned as uncertainty.

The systematic uncertainty of the background due to muons from J/$\psi$ decays is estimated considering the measured spectra at forward rapidity in Pb--Pb collisions at $\sNN$ = 5.02 TeV and their extrapolation to a larger $\pT$ and rapidity range. It varies within 4 and 8\%, depending on $\pT$, in central collisions. 

The systematic uncertainty of the normalisation factor, $F_{\rm norm}$, of 1\% is obtained by comparing the results obtained either using the relative trigger rates or applying the muon trigger condition in the analysis of MB events~\cite{Acharya:2018jvc}, see also section~\ref{sec:mudata}. 

The various systematic uncertainties are propagated to the measurement of muons from heavy-flavour hadron decays and added in quadrature\footnote{The systematic  uncertainty on $F_{\rm norm}$ is shown separately.}. The resulting systematic uncertainty decreases with increasing $\pT$ from about 5.5\% ($\pT$ = 3 GeV/$c$) to 3\% ($\pT$ = 8 GeV/$c$) in the 0--10\% centrality class and it varies within the range 4--6\% in peripheral collisions (40--60\% centrality interval).

In the electron analysis the systematic uncertainties are evaluated considering the following contributions: i) subtraction of electrons originating from Dalitz decays and photon conversions, including variation of the number of hits for electron candidates in the SPD, ii) matching efficiency of tracks reconstructed in the ITS and TPC, iii) matching of reconstructed tracks between the TOF and TPC, iv) track-reconstruction and identification procedure, v) space-charge distortions in the TPC drift volume, vi) residual hadron contamination, and vii) electron contribution from J/$\psi$ decays.

The main source of systematic uncertainty for $\pT~<~2~\text{GeV}/c$, where the signal to background ratio is of the order of a few percent, is related to the subtraction of electrons originating from Dalitz decays and photon conversions. This contribution is estimated as the RMS of the distribution of yields obtained by varying the selection criteria related to the associated particle. This systematic uncertainty has a maximum of about 25\% in the interval $0.2 < \pT < 0.3$ GeV/$c$ ($0.3 < \pT < 0.5$ GeV/$c$) in the 20--40\% (0--20\%) centrality class and it decreases to 3\% at high $\pT$. 
In order to further test the robustness of the background electron tagging, the requirement on the number of hits for electron candidates in the SPD is relaxed to increase the fraction of electrons from photon conversions in the detector material. A variation of 5\%, assigned as systematic uncertainty, is observed for the measured production yield up to $\pT$ = 2 GeV/$c$ for both centrality intervals, when one SPD hit is required for each track.
For $\pT~>$ 3 GeV/$c$, the uncertainties originating from the incomplete knowledge of the matching efficiency of tracks reconstructed in the ITS and TPC and between the TOF and TPC detectors
amount to 4\% and 2\%, respectively.
The systematic uncertainty introduced by the track-reconstruction procedure is estimated by varying the tracking parameters, like the number of TPC space-points belonging to a track, and it amounts to 4\%.
The systematic uncertainty introduced by the particle identification is also evaluated varying the electron selection criteria and is found to be negligible.
The effects due to the presence of non-uniformities in the correction for the space-charge distortion in the TPC drift volume or irregularities in the detector coverage are evaluated by repeating the analysis in different pseudorapidity regions. This gives a maximum systematic uncertainty of 10\% in the interval $0.2~<~\pT~<~0.3$ GeV/$c$. This contribution decreases to 4\% at $\pT$ = 0.4 GeV/$c$ and it is negligible for $\pT~>~2$~GeV/$c$.
The systematic uncertainty from the hadron contamination subtraction is estimated using different functional forms to fit the $n^{\rm TPC}_{\sigma}$ distribution and amounts to 6\% for the most central collisions and to 3\% in the 20--40\% centrality interval is assigned as in a previous publication \cite{Acharya:2019mom}.
Due to the non-subtraction of electrons from the J/$\psi$ decays, an uncertainty of 2\% is assigned for $\pT$ $<$ 3 GeV/$c$. This contribution increases to 4\% for $\pT$ $>$ 3 GeV/$c$ due to the growing contribution from J/$\psi$ decays observed in previous publications \cite{Acharya:2018upq, Acharya:2019mom}.
The overall systematic uncertainty of the $p_\mathrm{T}$-differential production yield is calculated summing in quadrature the different contributions.

\section{Derivation of production cross section of muons and electrons from heavy-flavour hadron decays in pp collisions at $\sqrt{s}$ = 5.44 TeV}

In order to calculate the nuclear modification factor of muons and electrons from heavy-flavour hadron decays, a reference production cross section for pp collisions at the same centre-of-mass energy as Xe--Xe collisions is needed for both analyses. Since pp data at $\sqrt{s}=5.44\text{ TeV}$ are currently not available, the pp reference is obtained by applying a pQCD-driven $\sqrt{s}$-scaling~\cite{Averbeck:2011ga} to the measured pp production cross section at $\sqrt{s}=5.02\text{ TeV}$ \cite{Acharya:2019mky,Acharya:2019mom}.
The $\pT$-dependent scaling factors are obtained by calculating the ratio of the production cross sections of electrons and muons from heavy-flavour hadron decays from FONLL calculations \cite{Cacciari:1998it} at $\sqrt{s}=5.5\text{ TeV}$\footnote{FONLL calculations are available at $\sqrt{s}$ = 5.5 TeV only. The difference with respect to $\sqrt{s}$ = 5.44 TeV is neglected.} to those at 
$\sqrt{s}=5.02\text{ TeV}$. 
The systematic uncertainty of the pp reference has two sources: the measured pp reference at $\sqrt s$ = 5.02 TeV and the $\pT$-dependent scaling factor from $\sqrt s$ = 5.02 TeV to $\sqrt s$ = 5.5 TeV. The former varies between about 3.4\% ($\pT$ = 8 GeV/$c$) to 4.2\% ($\pT$ = 3 GeV/$c$) for the muon case~\cite{Acharya:2019mky}, while for the electron case it varies between 10\% ($\pT$ = 0.5 GeV/$c$) and 5\% ($\pT$ = 6 GeV/$c$)~\cite{Acharya:2019mom}. The latter is in the range 1--1.5\% and includes the uncertainties on the parton distribution functions, quark masses, and factorisation and renormalisation scales, as described in~\cite{Averbeck:2011ga}. The two contributions are added in quadrature. In addition, the global normalisation uncertainty of 2.1\% evaluated from the pp analysis at $\sqrt s$ = 5.02 TeV is applied~\cite{Acharya:2019mky,ALICE-PUBLIC-2018-014,ALICE-PUBLIC-2016-005}. 

For the electron analysis, the pp production cross section is also extrapolated in $\pT$ since the pp production cross section at $\sqrt{s}=5.02\text{ TeV}$ was only measured for $\pT~>$ 0.5 GeV/$c$~\cite{Acharya:2019mom}, while in Xe--Xe collisions the electrons from heavy-flavour hadron decays are measured for $\pT~>$ 0.2 GeV/$c$. The production cross section obtained from FONLL calculations at $\sqrt{s}=5.5\text{ TeV}$ is used as the pp reference for $\pT < 0.5$ GeV/$c$. Since the central values
of the FONLL calculations underestimate the measurements of electrons from heavy-flavour hadron decays \cite{Acharya:2018upq, ALICE-PUBLIC-2018-005, Abelev:2012xe}, the FONLL production cross section is multiplied by a scaling factor, determined by fitting the data to theory ratio at $\sqrt s$ = 5.02 TeV with a second order polynomial function in the full $\pT$ range available for the measurement~\cite{Acharya:2019mom}. The systematic uncertainties for $\pT~<$ 0.5 GeV/$c$ are evaluated under the assumption that the systematic uncertainties of the measurement are fully correlated
over $\pT$, i.e. by repeating the calculation of the scaling factor after shifting all data points consistently within their systematic uncertainties. The resulting systematic uncertainty is 7\%.
An additional systematic uncertainty contribution of 20\% is assigned to account for the difference between the results from a first-order, third-order and fourth-order polynomial fit to the data-to-theory ratio and those from the second-order polynomial fit used for the central value.
Those two contributions are summed in quadrature to obtain the total systematic uncertainties.

\section{Results} 

\begin{figure}[ht!]
\begin{minipage}[c]{0.5\linewidth}
\includegraphics[width=1\textwidth]{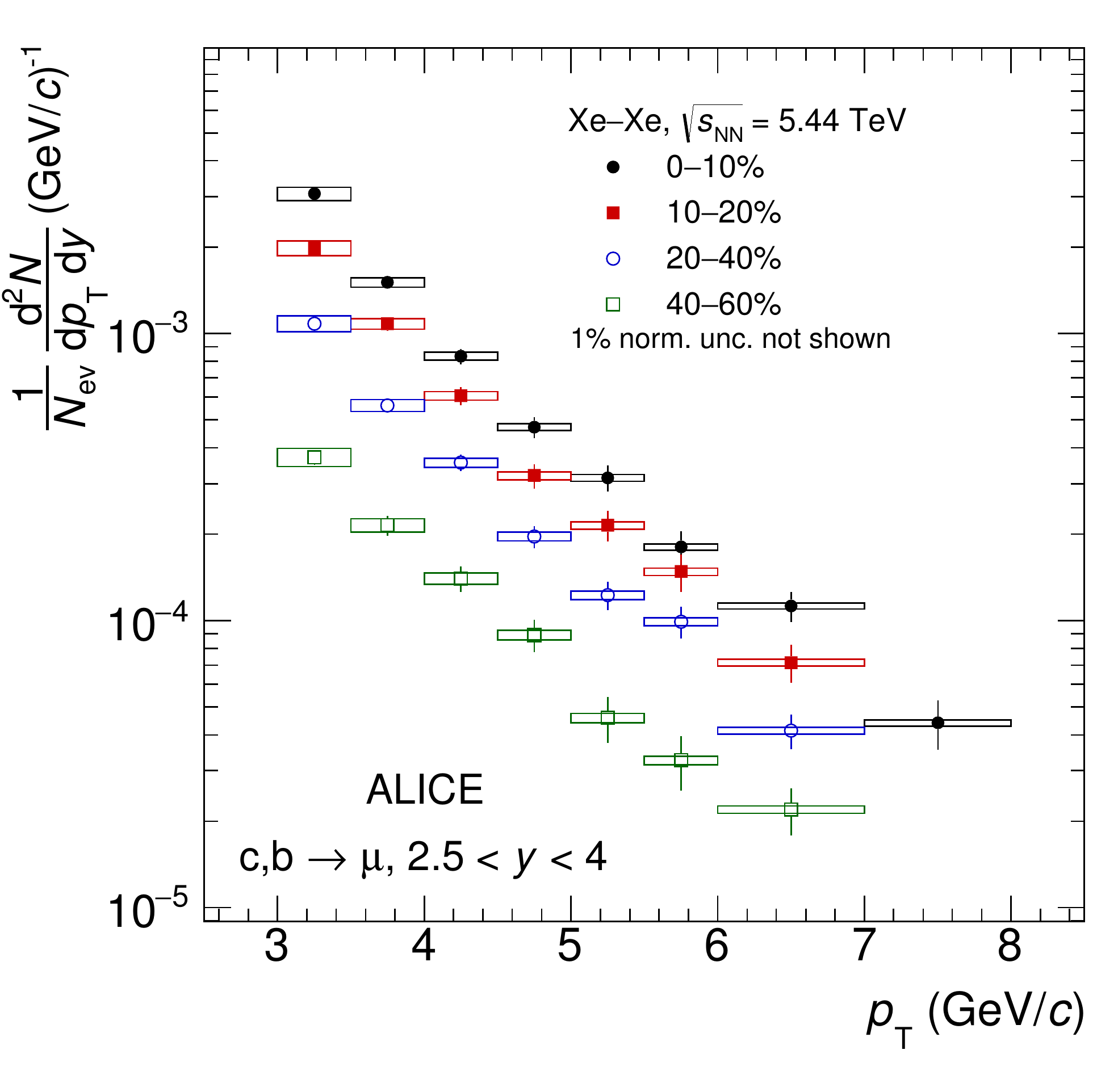}
\end{minipage}
\hspace{0.1cm}
\begin{minipage}[c]{0.5\linewidth}
\includegraphics[width=.9\textwidth]{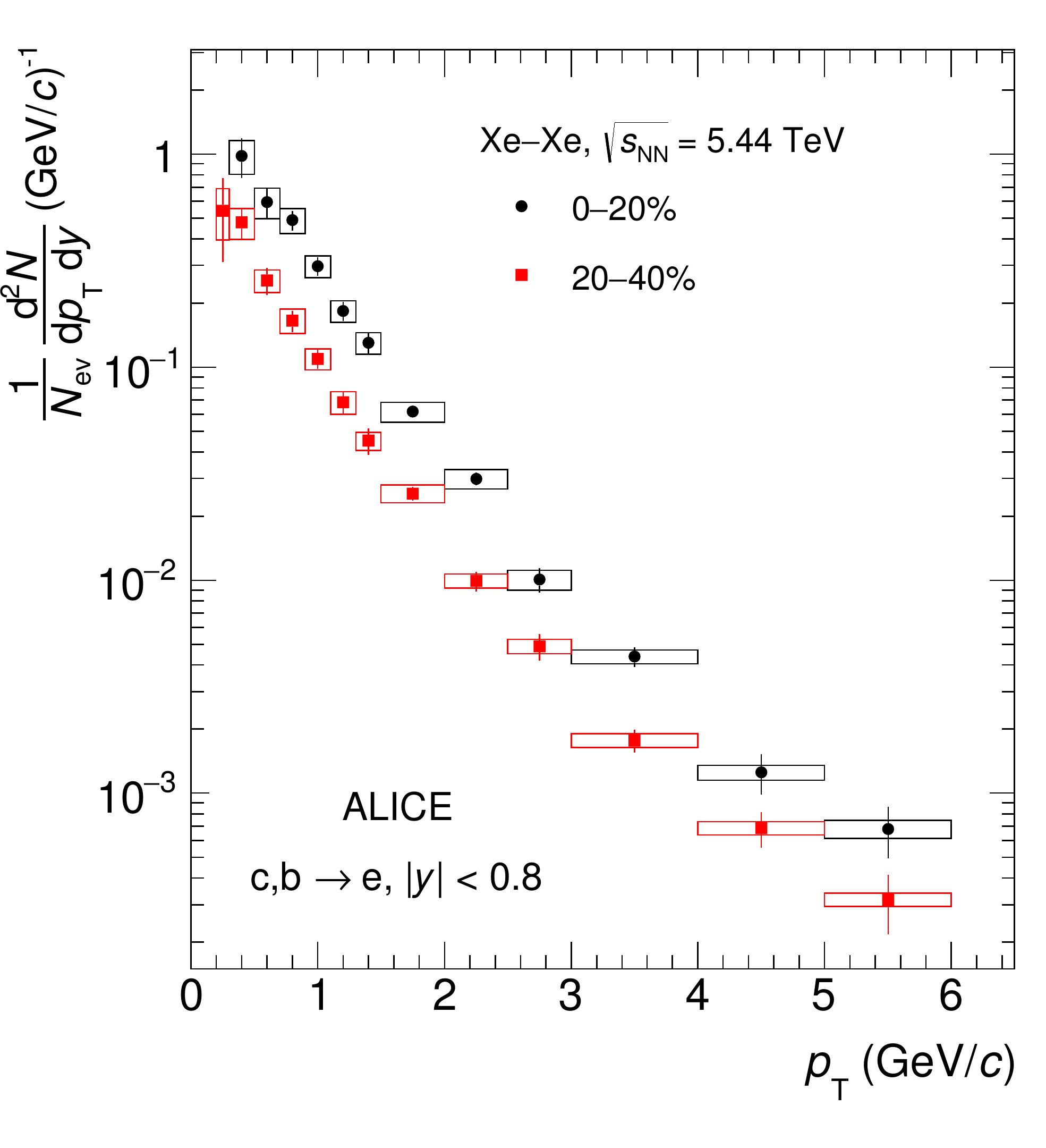}
\end{minipage}%
\caption{The $\pT$-differential production yields of muons (left panel) and electrons (right panel) from heavy-flavour hadron decays at forward and midrapidity, respectively, in Xe--Xe collisions at $\sqrt{s_{\rm NN}}$ = 5.44 TeV for various centrality intervals. Statistical uncertainties (vertical bars) and systematic uncertainties (open boxes) are shown.}
\label{fig:resultsspectrahfe}
\end{figure}

The $\pT$-differential production yields of muons and electrons from heavy-flavour hadron decays at forward and midrapidity in Xe--Xe collisions at $\sNN$ = 5.44 TeV are shown in the left and right panel of Fig.~\ref{fig:resultsspectrahfe}. 
The $\pT$-differential production yields are derived from the corrected and normalised differential yields of muons and electrons from heavy-flavour hadron decays, discussed in section~\ref{sec3}, which are further divided by a factor two introduced to obtain the charge-averaged differential yields. 

The muons from heavy-flavour hadron decays are measured in the centrality intervals 0--10\%, 10--20\%, 20--40\% and 40-60\% for the interval 3~$< \pT <$~7--8~GeV/$c$, while the electrons from heavy-flavour hadron decays are measured at midrapidity in the interval 0.2--0.3~$< \pT <$~6~GeV/$c$  in the 0--20\% and 20--40\% centrality intervals. The vertical bars denote  the statistical uncertainties and the systematic uncertainties are shown as empty boxes, except for the one on the normalisation factor, $F_{\rm norm}$, needed to determine the corresponding number of MB events in the muon-triggered sample (1\%) which is reported separately in the legend (left panel). 

\begin{figure}[ht!]
\centering
\includegraphics[width=.85\textwidth]{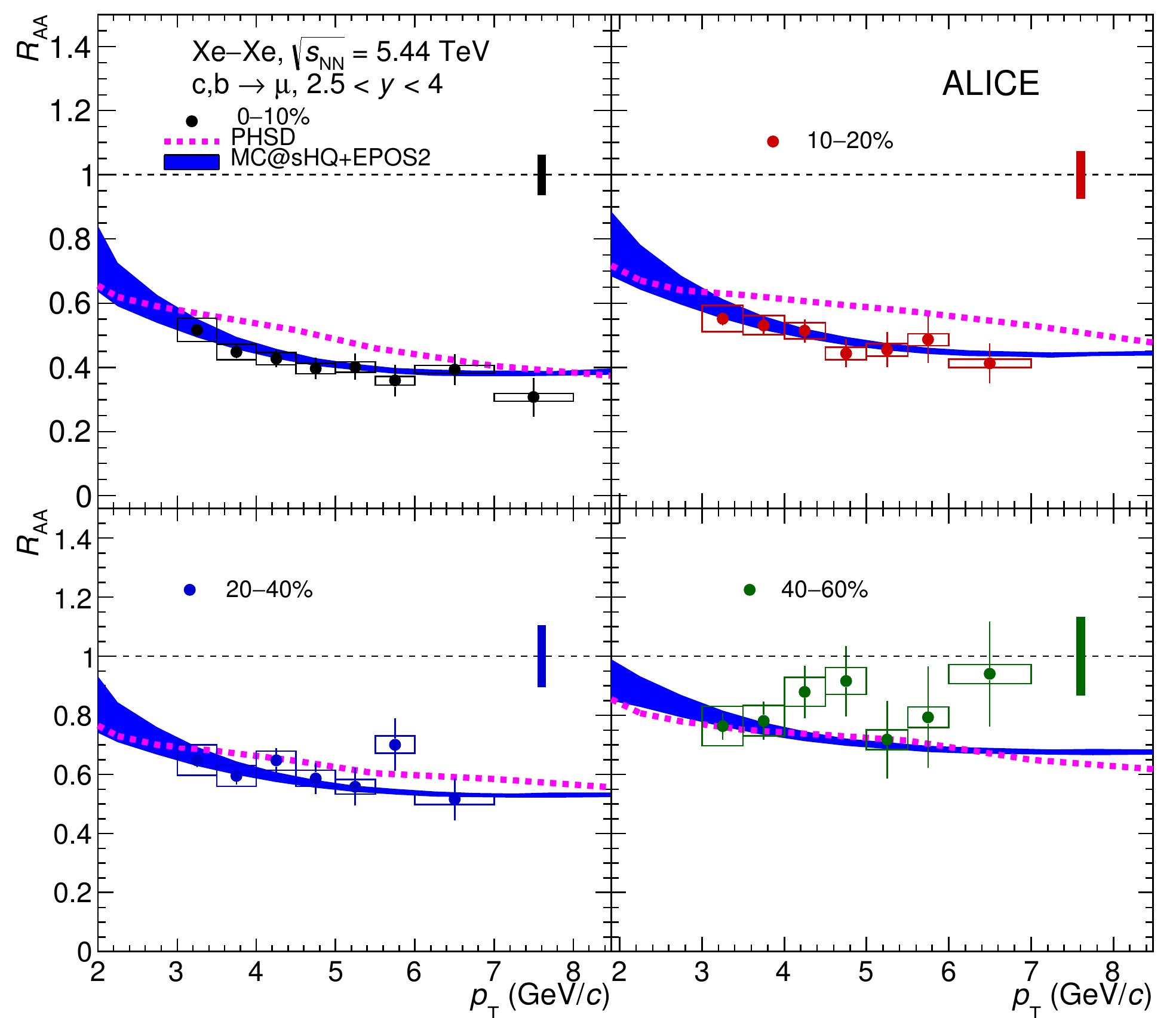}
\caption{Nuclear modification factor $\RAA$ of muons from heavy-flavour hadron decays at forward rapidity as a function of $\pT$ in Xe--Xe collisions at
$\sqrt{s_{\rm NN}}$ = 5.44 TeV for various centrality intervals mentioned in the figure. Statistical uncertainties (vertical bars) and systematic uncertainties 
(open boxes) are shown. The filled boxes at $\RAA$ = 1 represent the normalisation uncertainty. Comparisons with the PHSD~\cite{Song:2015ykw} and MC@sHQ+EPOS2~\cite{Nahrgang:2013xaa} models are presented. }
\label{fig:resultsraamuon}
\end{figure}

\begin{figure}[ht!]
\centering
\includegraphics[width=.9\textwidth]{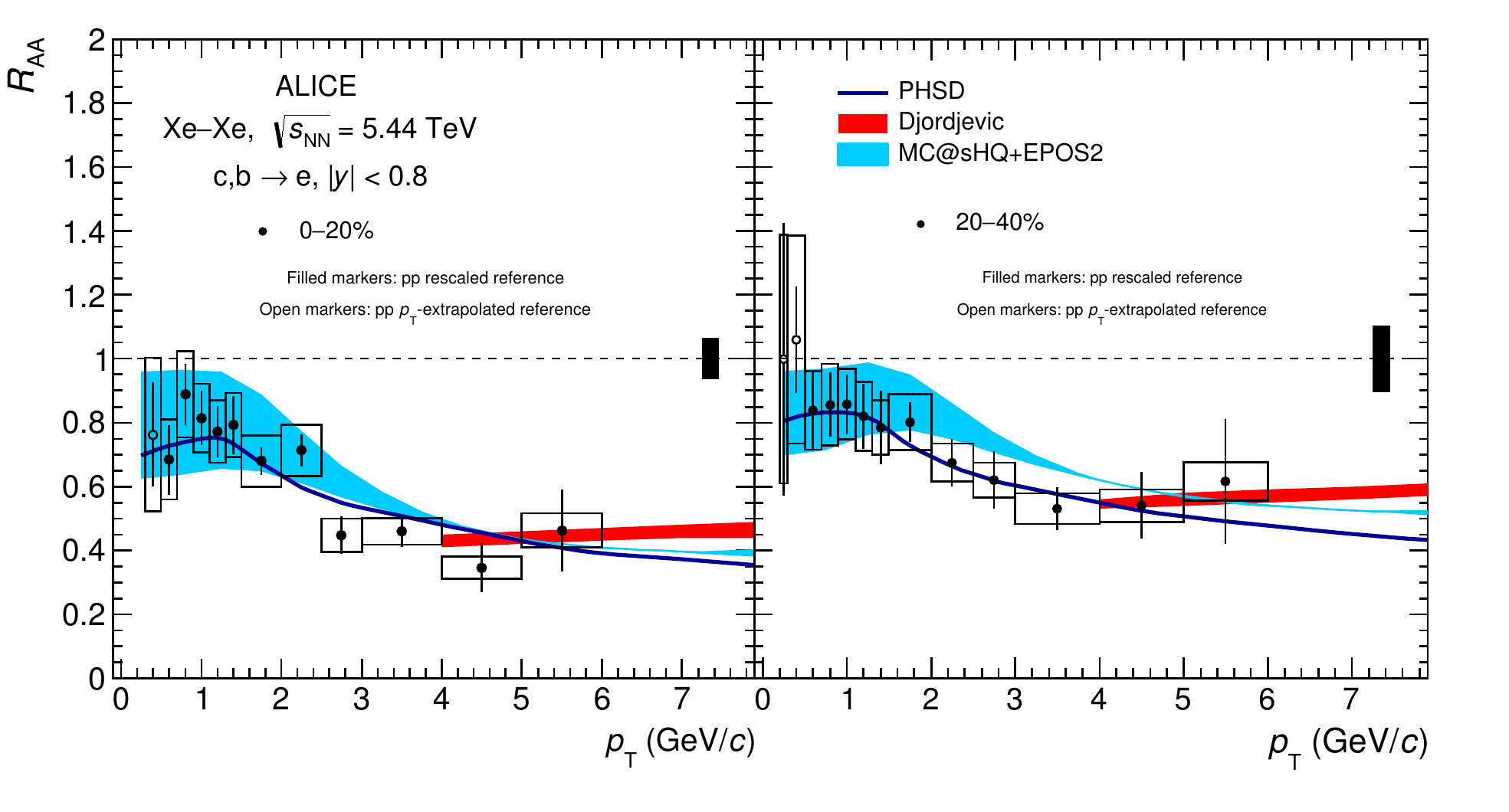}
\caption{Nuclear modification factor of electrons from semileptonic heavy-flavour hadron decays as a function of $\pT$ in Xe--Xe collisions at $\sqrt{s_{\rm NN}}$ = 5.44 TeV for the 0--20\% (left panel) and 20--40\% (right panel) centrality intervals.  Statistical uncertainties (vertical bars) and systematic uncertainties (open boxes) are shown. The filled boxes at $\RAA$ = 1 represent the normalisation uncertainty. Comparisons with the PHSD~\cite{Song:2015ykw}, MC@sHQ+EPOS2~\cite{Nahrgang:2013xaa} and Djordjevic~\cite{Djordjevic:2015hra, Djordjevic:2014} models are displayed.}
\label{fig:resultsraa}
\end{figure}

Figure~\ref{fig:resultsraamuon} displays the $\RAA$ of muons from heavy-flavour hadron decays as a function of $\pT$ in Xe--Xe collisions at $\sqrt{s_{\rm NN}}$~=~5.44~TeV for the centrality intervals 0--10\%, 10--20\%, 20--40\% and 40--60\%. The $\RAA$ of electrons from heavy-flavour hadron decays as a function of $\pT$ in Xe--Xe collisions at $\sqrt{s_{\rm NN}}$~=~5.44~TeV for the 0--20\% (left panel) and 20--40\% (right panel) centrality intervals is shown in Fig.~\ref{fig:resultsraa}.
For both figures, statistical (vertical bars) and systematic (empty boxes) uncertainties of the $\pT$-differential yields in Xe--Xe collisions and of the $\pT$-differential production cross section in pp collisions are propagated as uncorrelated uncertainties. The systematic uncertainty on the normalisation is indicated as a full box 
at $\RAA$ = 1. The latter is the quadratic sum of the systematic uncertainty of i) the average nuclear overlap function (Table \ref{tab::num_evs},~\cite{ALICE-PUBLIC-2018-011}), ii) the normalisation uncertainty of the pp reference and iii) the normalisation factor of muon-triggered events to the equivalent number of MB events (for the muon analysis only).

It is important to stress that the results at forward and central rapidity complement each other. 
In particular, the suppression pattern is similar for both rapidity regions, indicating that heavy quarks strongly interact with the medium created in heavy-ion collisions over a wide rapidity interval.

In the most central events, 
the $\RAA$ reaches its minimum of about 0.4 for $p_\mathrm{T} > 4$ GeV/$c$, while moving to more peripheral Xe--Xe collisions the $R_\mathrm{AA}$ gets closer to unity.
In order to better understand the origin of the observed suppression, it is interesting to note that measurements in minimum bias p-Pb collisions at $\sqrt{s_\textrm{NN}}$~=~5.02~TeV, where the formation of an extended QGP phase is not expected, show a nuclear modification factor compatible with unity at high $\pT$~\cite{Acharya:2017hdv,Adam:2015qda}. The comparison between the two systems confirms that the high-$\pT$ suppression of electrons and muons from heavy-flavour hadron decays is due to final-state effects, such as partonic energy loss in the medium, similar to the observation in Pb--Pb collisions~\cite{Acharya:2019mom}.
The centrality dependence of the $R_\textrm{AA}$ is compatible with the hypothesis that in-medium parton energy loss depends on the medium density. 
The latter increases towards most central collisions and thus the amount of energy lost by the parton traversing the medium increases.

For $\pT~<$~3~GeV/$c$, the $\RAA$ of electrons from heavy-flavour hadron decays is observed to increase with decreasing $\pT$. Such an increase is expected as it compensates for the suppression at higher $\pT$, assuming the scaling of the total yield with the number of binary nucleon-nucleon collisions in heavy-ion collisions~\cite{Radiativeb,Acharya:2019mom}.
However, the scaling can be broken due to the nuclear modification of the parton distribution functions in Xe nuclei~\cite{Eskola:2009uj}, leading to $\RAA$ values lower than unity also at low $\pT$. Transport model calculations, in order to better describe the low $\pT$ measurements, have to include shadowing effects, which reduce the $R_\textrm{AA}$ by about 30--40\% in the interval $\pT$ $<$ 5 GeV/$c$~\cite{Acharya:2018upq,Adam:2015sza}. 
In addition, further modifications of the $\pT$ distribution due to the radial flow can also play a role in this region \cite{PhysRevD.11.3105}. 

The $\RAA$ of muons and electrons from heavy-flavour hadron decays is compared with predictions from the PHSD and MC@sHQ+EPOS2 calculations~\cite{Song:2015ykw, Nahrgang:2013xaa}. The PHSD model considers the nuclear modification of the parton distribution functions and includes only collisional energy loss processes, while the MC@sHQ+EPOS2 model includes also energy loss from medium-induced gluon radiation. The MC@sHQ+EPOS2 calculations does not implement the nuclear modification of PDF for b quarks. Moreover, a contribution of hadronisation via recombination is considered in addition to the fragmentation mechanism. 
The MC@sHQ+EPOS2 calculations are displayed with their theoretical uncertainty band evaluated considering pure elastic and elastic+radiative energy loss together with the uncertainty on shadowing. 
The MC@sHQ+EPOS2 model provides a fair description of the $R_\mathrm{AA}$ of muons and electrons from heavy-flavour hadron decays within the experimental uncertainties in the full measured $\pT$ interval in all centrality intervals. The PHSD model is consistent with the data at central rapidity, while at forward rapidity the model has difficulties to reproduce the centrality dependence of the measured $\RAA$ and tends to overestimate the measured $\RAA$ in central collisions (0--10\% and 10--20\% centrality classes). The Djordjevic model \cite{Djordjevic:2015hra, Djordjevic:2014}, which implements energy loss for gluons, light and heavy quarks, including both radiative and collisional
processes and considering dynamical scattering centres in the medium, provides a good description of the $R_\mathrm{AA}$ of electrons from heavy-flavour hadron decays within uncertainties in both centrality intervals for $p_\mathrm{T}~>~4$~GeV/$c$.

\begin{figure}[!hbt]
\begin{center}
\includegraphics[width=.8\columnwidth]{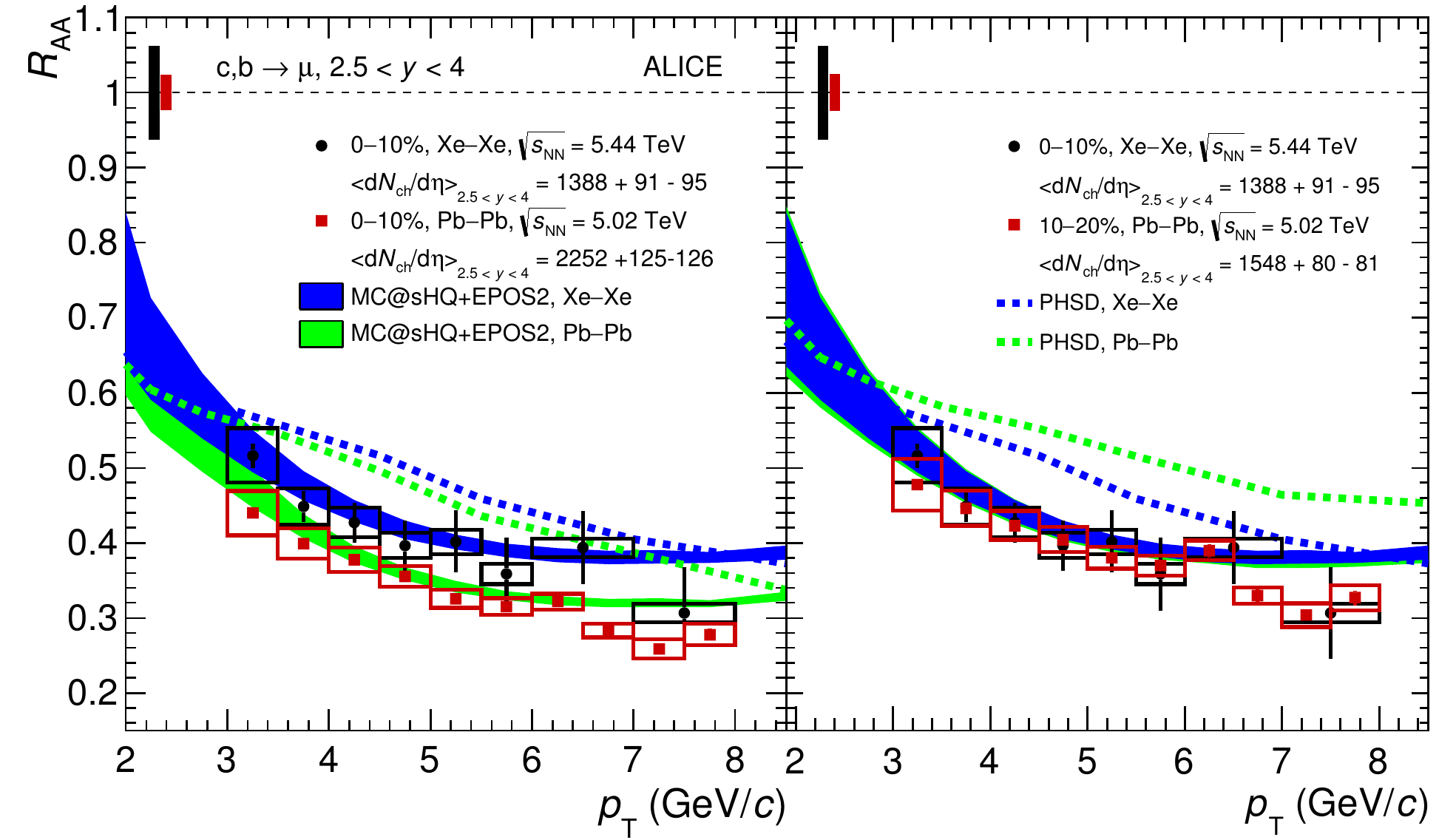}
\caption{Comparison of the $\pT$-differential nuclear modification factor of muons from heavy-flavour hadron decays at forward rapidity in Xe--Xe collisions at $\sNN$ = 5.44 TeV (centrality class 0--10\%) and Pb--Pb collisions at $\sNN$ = 5.02 TeV for the 
0--10\% (left) and 10--20\% (right) centrality classes. Statistical uncertainties (vertical bars) and systematic uncertainties (open boxes) are shown. The filled boxes at $\RAA$ = 1 represent the normalisation uncertainty. Comparisons with PHSD~\cite{Song:2015ykw} and MC@sHQ+EPOS2~\cite{Nahrgang:2013xaa} model predictions are also displayed.}
\label{Fig:pTdiffRAAPbXe}
\end{center}
\end{figure}

In a simple approach, the average energy loss depends on the density of scattering centres, which is proportional to the energy density, and on the path length of partons in the medium~\cite{dEnterria:2009xfs}. The energy density can be estimated from the average charged-particle multiplicity density per transverse area~\cite{Chatrchyan:2012mb,Acharya:2018hhy}. The path-length dependence of the energy loss is predicted to be linear for elastic (collisional) processes and quadratic for inelastic (radiative) processes~\cite{Baier:1996kr}. Consequently, the study of the system-size dependence of the production of leptons from heavy-flavour hadron decays is an important tool to investigate the path-length dependence of the in-medium parton energy loss in a hot and dense medium and to distinguish between different energy loss mechanisms~\cite{Djordjevic:2018ita}.

The nuclear modification factor of muons from heavy-flavour hadron decays in the 10\% most central Xe--Xe collisions at $\sqrt{s_{\rm NN}}$ = 5.44 TeV is compared in Fig.~\ref{Fig:pTdiffRAAPbXe} with that measured in Pb--Pb collisions at  $\sqrt{s_{\rm NN}}$ = 5.02 TeV for the 0--10\% and 10--20\% centrality classes~\cite{ALICE-PUBLIC-2020-009,pubPbPb} in the left and right panel, respectively. Comparisons with the PHSD and MC@sHQ+EPOS2 transport models~\cite{Song:2015ykw,Nahrgang:2013xaa} are also shown. In both collision systems, the $\pT$-differential $\RAA$ of muons from heavy-flavour hadron decays in central Xe--Xe collisions at $\sNN$ = 5.44 TeV shows a similar evolution as a function of $\pT$ when compared to the measured $\RAA$ in Pb--Pb collisions at $\sqrt{s_{\rm NN}}$ = 5.02 TeV for the same centrality class~\cite{pubPbPb}. 
However, a systematic difference between the two sets of $\RAA$ results is visible, which tends to indicate that the suppression is stronger in Pb--Pb collisions for the same centrality class. The \pT-integrated $R_{\rm AA}$ values, in the range 3-8 GeV/$c$, differ by about 2.5 standard deviations.
Such a behaviour may result from the difference in the system size. Both the MC@sHQ+EPOS2 and PHSD models predict a slightly larger suppression in Pb--Pb than in Xe--Xe collisions, as observed in the data. The MC@sHQ+EPOS2 model describes the suppression seen in the data for both central (0--10\%) Pb--Pb and Xe--Xe collisions. The PHSD calculations underestimate the measured suppression over the entire $\pT$ interval in these two collision systems. 

When the $\RAA$ measured in Xe--Xe and Pb--Pb collisions are compared for classes of events associated with a similar average charged-particle multiplicity density~\cite{Adam:2015ptt,Acharya:2018hhy}, such as 0--10\% for Xe--Xe collisions and 10--20\% for Pb--Pb collisions~\cite{pubPbPb,ALICE-PUBLIC-2020-009}, as in the right panel of Fig.~\ref{Fig:pTdiffRAAPbXe}, a remarkable agreement between the two collision systems is found. A similar effect was reported for high-$\pT$ charged particles measured at midrapidity by the ALICE collaboration~\cite{Acharya:2018eaq}. These similarities between the $\RAA$ of muons from heavy-flavour hadron decays in Xe--Xe and Pb--Pb collisions at comparable average charged-particle multiplicity density are in agreement with results from the study of the fractional momentum loss of high-$\pT$ partons at the RHIC and LHC~\cite{Adare:2015cua}.  While the MC@sHQ+EPOS2 calculations show a close similarity of the suppression pattern for Pb--Pb and Xe--Xe collisions over the whole $\pT$ interval, the PHSD model, with only collisional energy loss processes implemented, predicts here a larger suppression in Xe--Xe collisions compared to Pb--Pb collisions, in particular at high $\pT$. The MC@sHQ+EPOS2 calculations are in fair agreement with the measured $\RAA$ over the whole $\pT$ interval, both for Xe--Xe and Pb--Pb collisions. 
As already reported, the PHSD model tends to overestimate the measured $\RAA$ in central (0--10\% centrality class) Xe--Xe collisions. In central (10--20\% centrality class) Pb--Pb collisions, the PHSD calculations also systematically underestimate the measured suppression over the whole $\pT$ interval. Therefore, the study of the production of muons from heavy-flavour hadron decays at forward rapidity provides new constraints for the treatment of the path-length dependence of the different parton energy loss mechanisms in transport model calculations. 

\section{Summary and conclusions}

In summary, the first measurements of the nuclear modification factor $\RAA$ of muons and electrons from heavy-flavour hadron decays in Xe--Xe collisions at $\sNN$ = 5.44 TeV, performed with the ALICE detector at the LHC, are presented.
A strong suppression, reaching a factor of about 2.5 in central collisions, is observed at both central and forward rapidity. The study of Xe--Xe collisions provides an opportunity to further test the evolution of the in-medium parton energy loss with the system size. The $\RAA$ results in central Xe--Xe collisions show a smaller suppression than in Pb--Pb collisions when the measurements are performed in the same centrality class. In contrast, the measured $\RAA$ in central Xe--Xe collisions at forward rapidity is in agreement with that in Pb--Pb collisions, when these are compared for classes of events with similar charged-particle multiplicity density. The comparison of the $\RAA$ in the two collision systems as well as the $\RAA$ measurements in Xe--Xe collisions at central and forward rapidity for different centrality classes can shed more light on the path-length dependence of the in-medium parton energy loss.

\newenvironment{acknowledgement}{\relax}{\relax}
\begin{acknowledgement}
\section*{Acknowledgements}

The ALICE Collaboration would like to thank all its engineers and technicians for their invaluable contributions to the construction of the experiment and the CERN accelerator teams for the outstanding performance of the LHC complex.
The ALICE Collaboration gratefully acknowledges the resources and support provided by all Grid centres and the Worldwide LHC Computing Grid (WLCG) collaboration.
The ALICE Collaboration acknowledges the following funding agencies for their support in building and running the ALICE detector:
A. I. Alikhanyan National Science Laboratory (Yerevan Physics Institute) Foundation (ANSL), State Committee of Science and World Federation of Scientists (WFS), Armenia;
Austrian Academy of Sciences, Austrian Science Fund (FWF): [M 2467-N36] and Nationalstiftung f\"{u}r Forschung, Technologie und Entwicklung, Austria;
Ministry of Communications and High Technologies, National Nuclear Research Center, Azerbaijan;
Conselho Nacional de Desenvolvimento Cient\'{\i}fico e Tecnol\'{o}gico (CNPq), Financiadora de Estudos e Projetos (Finep), Funda\c{c}\~{a}o de Amparo \`{a} Pesquisa do Estado de S\~{a}o Paulo (FAPESP) and Universidade Federal do Rio Grande do Sul (UFRGS), Brazil;
Ministry of Education of China (MOEC) , Ministry of Science \& Technology of China (MSTC) and National Natural Science Foundation of China (NSFC), China;
Ministry of Science and Education and Croatian Science Foundation, Croatia;
Centro de Aplicaciones Tecnol\'{o}gicas y Desarrollo Nuclear (CEADEN), Cubaenerg\'{\i}a, Cuba;
Ministry of Education, Youth and Sports of the Czech Republic, Czech Republic;
The Danish Council for Independent Research | Natural Sciences, the VILLUM FONDEN and Danish National Research Foundation (DNRF), Denmark;
Helsinki Institute of Physics (HIP), Finland;
Commissariat \`{a} l'Energie Atomique (CEA) and Institut National de Physique Nucl\'{e}aire et de Physique des Particules (IN2P3) and Centre National de la Recherche Scientifique (CNRS), France;
Bundesministerium f\"{u}r Bildung und Forschung (BMBF) and GSI Helmholtzzentrum f\"{u}r Schwerionenforschung GmbH, Germany;
General Secretariat for Research and Technology, Ministry of Education, Research and Religions, Greece;
National Research, Development and Innovation Office, Hungary;
Department of Atomic Energy Government of India (DAE), Department of Science and Technology, Government of India (DST), University Grants Commission, Government of India (UGC) and Council of Scientific and Industrial Research (CSIR), India;
Indonesian Institute of Science, Indonesia;
Istituto Nazionale di Fisica Nucleare (INFN), Italy;
Institute for Innovative Science and Technology , Nagasaki Institute of Applied Science (IIST), Japanese Ministry of Education, Culture, Sports, Science and Technology (MEXT) and Japan Society for the Promotion of Science (JSPS) KAKENHI, Japan;
Consejo Nacional de Ciencia (CONACYT) y Tecnolog\'{i}a, through Fondo de Cooperaci\'{o}n Internacional en Ciencia y Tecnolog\'{i}a (FONCICYT) and Direcci\'{o}n General de Asuntos del Personal Academico (DGAPA), Mexico;
Nederlandse Organisatie voor Wetenschappelijk Onderzoek (NWO), Netherlands;
The Research Council of Norway, Norway;
Commission on Science and Technology for Sustainable Development in the South (COMSATS), Pakistan;
Pontificia Universidad Cat\'{o}lica del Per\'{u}, Peru;
Ministry of Science and Higher Education, National Science Centre and WUT ID-UB, Poland;
Korea Institute of Science and Technology Information and National Research Foundation of Korea (NRF), Republic of Korea;
Ministry of Education and Scientific Research, Institute of Atomic Physics and Ministry of Research and Innovation and Institute of Atomic Physics, Romania;
Joint Institute for Nuclear Research (JINR), Ministry of Education and Science of the Russian Federation, National Research Centre Kurchatov Institute, Russian Science Foundation and Russian Foundation for Basic Research, Russia;
Ministry of Education, Science, Research and Sport of the Slovak Republic, Slovakia;
National Research Foundation of South Africa, South Africa;
Swedish Research Council (VR) and Knut \& Alice Wallenberg Foundation (KAW), Sweden;
European Organization for Nuclear Research, Switzerland;
Suranaree University of Technology (SUT), National Science and Technology Development Agency (NSDTA) and Office of the Higher Education Commission under NRU project of Thailand, Thailand;
Turkish Atomic Energy Agency (TAEK), Turkey;
National Academy of  Sciences of Ukraine, Ukraine;
Science and Technology Facilities Council (STFC), United Kingdom;
National Science Foundation of the United States of America (NSF) and United States Department of Energy, Office of Nuclear Physics (DOE NP), United States of America. 
\end{acknowledgement}

\bibliographystyle{utphys} 
\bibliography{AliHFMEXeXe-new}

\providecommand{\href}[2]{#2}\begingroup\raggedright\begin{thebibliography}{10}

\bibitem{Muller:2006ee}
B.~Muller and J.~L. Nagle, ``{Results from the relativistic heavy ion
  collider}'',
  \href{http://dx.doi.org/10.1146/annurev.nucl.56.080805.140556}{{\em Ann. Rev.
  Nucl. Part. Sci.} {\bfseries 56} (2006) 93--135},
  \href{http://arxiv.org/abs/nucl-th/0602029}{{\ttfamily
  arXiv:nucl-th/0602029}}.

\bibitem{Bazavov:2011nk}
A.~Bazavov {\em et~al.}, ``{The chiral and deconfinement aspects of the QCD
  transition}'', \href{http://dx.doi.org/10.1103/PhysRevD.85.054503}{{\em Phys.
  Rev. D} {\bfseries 85} (2012) 054503},
  \href{http://arxiv.org/abs/1111.1710}{{\ttfamily arXiv:1111.1710 [hep-lat]}}.

\bibitem{pQCD2}
{\bfseries Wuppertal-Budapest} Collaboration, S.~Borsanyi, Z.~Fodor,
  C.~Hoelbling, S.~D. Katz, S.~Krieg, C.~Ratti, and K.~K. Szabo, ``{Is there
  still any $T_{c}$ mystery in lattice QCD? Results with physical masses in the
  continuum limit III}'', \href{http://dx.doi.org/10.1007/JHEP09(2010)073}{{\em
  JHEP} {\bfseries 09} (2010) 073},
\href{http://arxiv.org/abs/1005.3508}{{\ttfamily arXiv:1005.3508 [hep-lat]}}.

\bibitem{Bazavov:2014pvz}
{\bfseries HotQCD} Collaboration, A.~Bazavov {\em et~al.}, ``{Equation of state
  in ( 2+1 )-flavor QCD}'',
  \href{http://dx.doi.org/10.1103/PhysRevD.90.094503}{{\em Phys. Rev.}
  {\bfseries D90} (2014) 094503},
\href{http://arxiv.org/abs/1407.6387}{{\ttfamily arXiv:1407.6387 [hep-lat]}}.

\bibitem{Borsanyi:2013bia}
S.~Borsanyi, Z.~Fodor, C.~Hoelbling, S.~D. Katz, S.~Krieg, and K.~K. Szabo,
  ``{Full result for the QCD equation of state with 2+1 flavors}'',
  \href{http://dx.doi.org/10.1016/j.physletb.2014.01.007}{{\em Phys. Lett.}
  {\bfseries B730} (2014) 99},
\href{http://arxiv.org/abs/1309.5258}{{\ttfamily arXiv:1309.5258 [hep-lat]}}.

\bibitem{Zyla:2020zbs}
{\bfseries Particle Data Group} Collaboration, P.~Zyla {\em et~al.}, ``{Review
  of Particle Physics}'', \href{http://dx.doi.org/10.1093/ptep/ptaa104}{{\em
  PTEP} {\bfseries 2020} no.~8, (2020) 083C01}.

\bibitem{Liu:2012ax}
F.-M. Liu and S.-X. Liu, ``{Quark-gluon plasma formation time and direct
  photons from heavy ion collisions}'',
  \href{http://dx.doi.org/10.1103/PhysRevC.89.034906}{{\em Phys. Rev.}
  {\bfseries C89} (2014) 034906},
\href{http://arxiv.org/abs/1212.6587}{{\ttfamily arXiv:1212.6587 [nucl-th]}}.

\bibitem{bib::Averbeck}
R.~Averbeck, ``{Heavy-flavor production in heavy-ion collisions and
  implications for the properties of hot QCD matter}'',
  \href{http://dx.doi.org/10.1016/j.ppnp.2013.01.001}{{\em Prog. Part. Nucl.
  Phys.} {\bfseries 70} (2013) 159},
\href{http://arxiv.org/abs/1505.03828}{{\ttfamily arXiv:1505.03828 [nucl-ex]}}.

\bibitem{bib::Braun_Munziger}
P.~Braun-Munzinger, ``{Quarkonium production in ultra-relativistic nuclear
  collisions: Suppression versus enhancement}'',
  \href{http://dx.doi.org/10.1088/0954-3899/34/8/S36}{{\em J. Phys.} {\bfseries
  G34} (2007) S471},
\href{http://arxiv.org/abs/nucl-th/0701093}{{\ttfamily arXiv:nucl-th/0701093
  [NUCL-TH]}}.

\bibitem{intro1}
M.~Gyulassy and M.~Plumer, ``{Jet Quenching in Dense Matter}'',
\href{http://dx.doi.org/10.1016/0370-2693(90)91409-5}{{\em Phys. Lett.}
  {\bfseries B243} (1990) 432}.

\bibitem{Radiativeb}
R.~Baier, Y.~L. Dokshitzer, A.~H. Mueller, S.~Peigne, and D.~Schiff,
  ``{Radiative energy loss and $p_{\rm T}$-broadening of high-energy partons in
  nuclei}'', \href{http://dx.doi.org/10.1016/S0550-3213(96)00581-0}{{\em Nucl.
  Phys.} {\bfseries B484} (1997) 265},
\href{http://arxiv.org/abs/hep-ph/9608322}{{\ttfamily arXiv:hep-ph/9608322
  [hep-ph]}}.

\bibitem{Colla}
M.~H. Thoma and M.~Gyulassy, ``Quark damping and energy loss in the high
  temperature {QCD}'',
  \href{http://dx.doi.org/http://dx.doi.org/10.1016/S0550-3213(05)80031-8}{{\em
  Nucl.. Phys.} {\bfseries B351} (1991) 491}.
  \url{http://www.sciencedirect.com/science/article/pii/S0550321305800318}.

\bibitem{Collc}
E.~Braaten and M.~H. Thoma, ``Energy loss of a heavy quark in the quark-gluon
  plasma'', \href{http://dx.doi.org/10.1103/PhysRevD.44.R2625}{{\em Phys. Rev.
  D} {\bfseries 44} (1991) R2625}.
  \url{http://link.aps.org/doi/10.1103/PhysRevD.44.R2625}.

\bibitem{Glauber:1970jm}
R.~J. Glauber and G.~Matthiae, ``{High-energy scattering of protons by
  nuclei}'',
\href{http://dx.doi.org/10.1016/0550-3213(70)90511-0}{{\em Nucl. Phys.}
  {\bfseries B21} (1970) 135}.

\bibitem{Miller:2007ri}
M.~L. Miller, K.~Reygers, S.~J. Sanders, and P.~Steinberg, ``{Glauber modeling
  in high energy nuclear collisions}'',
  \href{http://dx.doi.org/10.1146/annurev.nucl.57.090506.123020}{{\em Ann. Rev.
  Nucl. Part. Sci.} {\bfseries 57} (2007) 205},
\href{http://arxiv.org/abs/nucl-ex/0701025}{{\ttfamily arXiv:nucl-ex/0701025
  [nucl-ex]}}.

\bibitem{ALICE-PUBLIC-2018-011}
{\bfseries ALICE} Collaboration, ``{Centrality determination in heavy ion
  collisions}.'' {ALICE-PUBLIC-2018-011}, 2018.
\newblock \url{http://cds.cern.ch/record/2636623}.

\bibitem{Andronic:2015wma}
A.~Andronic {\em et~al.}, ``{Heavy-flavour and quarkonium production in the LHC
  era: from proton--proton to heavy-ion collisions}'',
  \href{http://dx.doi.org/10.1140/epjc/s10052-015-3819-5}{{\em Eur. Phys. J. C}
  {\bfseries 76} (2016) 107}, \href{http://arxiv.org/abs/1506.03981}{{\ttfamily
  arXiv:1506.03981 [nucl-ex]}}.

\bibitem{Adam:2015nna}
{\bfseries ALICE} Collaboration, J.~Adam {\em et~al.}, ``{Centrality dependence
  of high-p$_{T}$ D meson suppression in Pb-Pb collisions at $
  \sqrt{s_{\mathrm{N}\mathrm{N}}}=2.76 $ TeV}'',
  \href{http://dx.doi.org/10.1007/JHEP11(2015)205,
  10.1007/JHEP06(2017)032}{{\em JHEP} {\bfseries 11} (2015) 205},
  \href{http://arxiv.org/abs/1506.06604}{{\ttfamily arXiv:1506.06604
  [nucl-ex]}}.
[Addendum: JHEP06,032(2017)].

\bibitem{Sirunyan:2017xss}
{\bfseries CMS} Collaboration, A.~M. Sirunyan {\em et~al.}, ``{Nuclear
  modification factor of D$^0$ mesons in PbPb collisions at
  $\sqrt{s_\mathrm{NN}} = 5.02$ TeV}'',
  \href{http://dx.doi.org/10.1016/j.physletb.2018.05.074}{{\em Phys. Lett.}
  {\bfseries B782} (2018) 474},
\href{http://arxiv.org/abs/1708.04962}{{\ttfamily arXiv:1708.04962 [nucl-ex]}}.

\bibitem{Acharya:2018hre}
{\bfseries ALICE} Collaboration, S.~Acharya {\em et~al.}, ``{Measurement of
  D$^{0}$, D$^{+}$, D$^{*+}$ and D$_{s}^{+}$ production in Pb--Pb collisions at
  $ \sqrt{{\mathrm{s}}_{\mathrm{NN}}}=5.02 $ TeV}'',
  \href{http://dx.doi.org/10.1007/JHEP10(2018)174}{{\em JHEP} {\bfseries 10}
  (2018) 174},
\href{http://arxiv.org/abs/1804.09083}{{\ttfamily arXiv:1804.09083 [nucl-ex]}}.

\bibitem{Sirunyan:2017oug}
{\bfseries CMS} Collaboration, A.~M. Sirunyan {\em et~al.}, ``{Measurement of
  the ${B}^{\pm}$ Meson Nuclear Modification Factor in Pb-Pb Collisions at
  $\sqrt{{s}_{\rm NN}}$ = 5.02 TeV}'',
  \href{http://dx.doi.org/10.1103/PhysRevLett.119.152301}{{\em Phys. Rev.
  Lett.} {\bfseries 119} no.~15, (2017) 152301},
\href{http://arxiv.org/abs/1705.04727}{{\ttfamily arXiv:1705.04727 [hep-ex]}}.

\bibitem{Aaboud:2018bdg}
{\bfseries ATLAS} Collaboration, M.~Aaboud {\em et~al.}, ``{Measurement of the
  suppression and azimuthal anisotropy of muons from heavy-flavor decays in
  Pb+Pb collisions at $\sqrt{s_{\mathrm{NN}}} = 2.76$ TeV with the ATLAS
  detector}'', \href{http://dx.doi.org/10.1103/PhysRevC.98.044905}{{\em Phys.
  Rev.} {\bfseries C98} (2018) 044905},
\href{http://arxiv.org/abs/1805.05220}{{\ttfamily arXiv:1805.05220 [nucl-ex]}}.

\bibitem{Acharya:2019mom}
{\bfseries ALICE} Collaboration, S.~Acharya {\em et~al.}, ``{Measurement of
  electrons from semileptonic heavy-flavour hadron decays at midrapidity in pp
  and Pb--Pb collisions at $\sqrt{s_{\rm{NN}}}$ = 5.02 TeV}'',
  \href{http://dx.doi.org/10.1016/j.physletb.2020.135377}{{\em Phys. Lett. B}
  {\bfseries 804} (2020) 135377},
  \href{http://arxiv.org/abs/1910.09110}{{\ttfamily arXiv:1910.09110
  [nucl-ex]}}.

\bibitem{pubPbPb}
{\bfseries ALICE} Collaboration, S.~Acharya {\em et~al.}, ``{Production of
  muons from heavy-flavour hadron decays at high transverse momentum in Pb-Pb
  collisions at $\sqrt{s_{\rm NN}}=5.02$ and 2.76 TeV}'',
  \href{http://arxiv.org/abs/2011.05718}{{\ttfamily arXiv:2011.05718
  [nucl-ex]}}.

\bibitem{Adam:2016wyz}
{\bfseries ALICE} Collaboration, J.~Adam {\em et~al.}, ``{Measurement of
  electrons from beauty-hadron decays in p-Pb collisions at $
  \sqrt{s_{\mathrm{NN}}}=5.02 $ TeV and Pb-Pb collisions at $
  \sqrt{s_{\mathrm{NN}}}=2.76 $ TeV}'',
  \href{http://dx.doi.org/10.1007/JHEP07(2017)052}{{\em JHEP} {\bfseries 07}
  (2017) 052},
\href{http://arxiv.org/abs/1609.03898}{{\ttfamily arXiv:1609.03898 [nucl-ex]}}.

\bibitem{Adam:2015qda}
{\bfseries ALICE} Collaboration, J.~Adam {\em et~al.}, ``{Measurement of
  electrons from heavy-flavour hadron decays in p--Pb collisions at
  $\sqrt{s_{\rm NN}} =$ 5.02 TeV}'',
  \href{http://dx.doi.org/10.1016/j.physletb.2015.12.067}{{\em Phys. Lett.}
  {\bfseries B754} (2016) 81},
\href{http://arxiv.org/abs/1509.07491}{{\ttfamily arXiv:1509.07491 [nucl-ex]}}.

\bibitem{Abelev:2014hha}
{\bfseries ALICE} Collaboration, B.~Abelev {\em et~al.}, ``{Measurement of
  prompt $D$-meson production in p--Pb collisions at $\sqrt{s_{\rm NN}}$ = 5.02
  TeV}'', \href{http://dx.doi.org/10.1103/PhysRevLett.113.232301}{{\em Phys.
  Rev. Lett.} {\bfseries 113} no.~23, (2014) 232301},
\href{http://arxiv.org/abs/1405.3452}{{\ttfamily arXiv:1405.3452 [nucl-ex]}}.

\bibitem{Eskola:2009uj}
K.~J. Eskola, H.~Paukkunen, and C.~A. Salgado, ``{EPS09: A New Generation of
  NLO and LO Nuclear Parton Distribution Functions}'',
  \href{http://dx.doi.org/10.1088/1126-6708/2009/04/065}{{\em JHEP} {\bfseries
  04} (2009) 065},
\href{http://arxiv.org/abs/0902.4154}{{\ttfamily arXiv:0902.4154 [hep-ph]}}.

\bibitem{Greco:2003vf}
V.~Greco, C.~M. Ko, and R.~Rapp, ``{Quark coalescence for charmed mesons in
  ultrarelativistic heavy ion collisions}'',
  \href{http://dx.doi.org/10.1016/j.physletb.2004.06.064}{{\em Phys. Lett.}
  {\bfseries B595} (2004) 202},
\href{http://arxiv.org/abs/nucl-th/0312100}{{\ttfamily arXiv:nucl-th/0312100
  [nucl-th]}}.

\bibitem{Djordjevic:2018ita}
M.~Djordjevic, D.~Zigic, M.~Djordjevic, and J.~Auvinen, ``{How to test
  path-length dependence in energy loss mechanisms: analysis leading to a new
  observable}'', \href{http://dx.doi.org/10.1103/PhysRevC.99.061902}{{\em Phys.
  Rev.} {\bfseries C99} no.~6, (2019) 061902},
\href{http://arxiv.org/abs/1805.04030}{{\ttfamily arXiv:1805.04030 [nucl-th]}}.

\bibitem{PHENIXRaamuCuCu}
{\bfseries PHENIX} Collaboration, A.~Adare {\em et~al.},
  ``{Nuclear-Modification Factor for Open-Heavy-Flavor Production at Forward
  Rapidity in Cu+Cu Collisions at $\sqrt{s_{NN}}=200$ GeV}'',
  \href{http://dx.doi.org/10.1103/PhysRevC.86.024909}{{\em Phys. Rev.}
  {\bfseries C86} (2012) 024909},
\href{http://arxiv.org/abs/1204.0754}{{\ttfamily arXiv:1204.0754 [nucl-ex]}}.

\bibitem{Adare:2013yxp}
{\bfseries PHENIX} Collaboration, A.~Adare {\em et~al.}, ``{System-size
  dependence of open-heavy-flavor production in nucleus-nucleus collisions at
  $\sqrt{s_{_{\rm NN}}}$ = 200 GeV}'',
  \href{http://dx.doi.org/10.1103/PhysRevC.90.034903}{{\em Phys. Rev.}
  {\bfseries C90} no.~3, (2014) 034903},
\href{http://arxiv.org/abs/1310.8286}{{\ttfamily arXiv:1310.8286 [nucl-ex]}}.

\bibitem{Sirunyan:2018eqi}
{\bfseries CMS} Collaboration, A.~M. Sirunyan {\em et~al.}, ``{Charged-particle
  nuclear modification factors in Xe-Xe collisions at $ \sqrt{s_{\mathrm{NN}}}
  = 5.44 $ TeV}'', \href{http://dx.doi.org/10.1007/JHEP10(2018)138}{{\em JHEP}
  {\bfseries 10} (2018) 138},
\href{http://arxiv.org/abs/1809.00201}{{\ttfamily arXiv:1809.00201 [hep-ex]}}.

\bibitem{Acharya:2018eaq}
{\bfseries ALICE} Collaboration, S.~Acharya {\em et~al.}, ``{Transverse
  momentum spectra and nuclear modification factors of charged particles in
  Xe--Xe collisions at $\sqrt{s_{\rm NN}}$ = 5.44 TeV}'',
  \href{http://dx.doi.org/10.1016/j.physletb.2018.10.052}{{\em Phys. Lett.}
  {\bfseries B788} (2019) 166},
\href{http://arxiv.org/abs/1805.04399}{{\ttfamily arXiv:1805.04399 [nucl-ex]}}.

\bibitem{Acharya:2018jvc}
{\bfseries ALICE} Collaboration, S.~Acharya {\em et~al.}, ``{Inclusive J/$\psi$
  production in Xe--Xe collisions at $\sqrt{s_{\rm NN}}$ = 5.44 TeV}'',
  \href{http://dx.doi.org/10.1016/j.physletb.2018.08.047}{{\em Phys. Lett.}
  {\bfseries B785} (2018) 419},
\href{http://arxiv.org/abs/1805.04383}{{\ttfamily arXiv:1805.04383 [nucl-ex]}}.

\bibitem{Aamodt:2008zz}
{\bfseries ALICE} Collaboration, K.~Aamodt {\em et~al.}, ``{The ALICE
  experiment at the CERN LHC}'',
\href{http://dx.doi.org/10.1088/1748-0221/3/08/S08002}{{\em JINST} {\bfseries
  3} (2008) S08002}.

\bibitem{bib::Alice_performance}
{\bfseries ALICE} Collaboration, B.~Abelev {\em et~al.}, ``{Performance of the
  ALICE Experiment at the CERN LHC}'',
  \href{http://dx.doi.org/10.1142/S0217751X14300440}{{\em Int.J.Mod.Phys.}
  {\bfseries A29} (2014) 1430044},
\href{http://arxiv.org/abs/1402.4476}{{\ttfamily arXiv:1402.4476 [nucl-ex]}}.

\bibitem{CERN-LHCC-2000-046}
{\bfseries ALICE} Collaboration, {\em {ALICE dimuon forward spectrometer:
  addendum to the Technical Design Report}}.
\newblock Technical Design Report ALICE. CERN, Geneva, 2000.
\newblock \url{https://cds.cern.ch/record/494265}.

\bibitem{Abelev:2013qoq}
{\bfseries ALICE} Collaboration, B.~Abelev {\em et~al.}, ``{Centrality
  determination of Pb--Pb collisions at $\sqrt{s_{NN}}$ = 2.76 TeV with
  ALICE}'', \href{http://dx.doi.org/10.1103/PhysRevC.88.044909}{{\em Phys.
  Rev.} {\bfseries C88} no.~4, (2013) 044909},
\href{http://arxiv.org/abs/1301.4361}{{\ttfamily arXiv:1301.4361 [nucl-ex]}}.

\bibitem{ALICEDRPbPbcent}
{\bfseries ALICE} Collaboration, J.~Adam {\em et~al.}, ``{Centrality dependence
  of high-$p_{\rm T}$ D meson suppression in {{Pb--Pb}} collisions at
  $\sqrt{s_\mathrm{NN}}$ = 2.76 {{TeV}}}'',
  \href{http://dx.doi.org/10.1007/JHEP11(2015)205}{{\em JHEP} {\bfseries 11}
  (2015) 205},
\href{http://arxiv.org/abs/1506.06604}{{\ttfamily arXiv:1506.06604 [nucl-ex]}}.

\bibitem{Loizides:2017ack}
C.~Loizides, J.~Kamin, and D.~d'Enterria, ``{Improved Monte Carlo Glauber
  predictions at present and future nuclear colliders}'',
  \href{http://dx.doi.org/10.1103/PhysRevC.97.054910,
  10.1103/PhysRevC.99.019901}{{\em Phys. Rev.} {\bfseries C97} (2018) 054910},
  \href{http://arxiv.org/abs/1710.07098}{{\ttfamily arXiv:1710.07098
  [nucl-ex]}}.
[erratum: Phys. Rev.C99,no.1,019901(2019)].

\bibitem{Aamodt:2011gj}
{\bfseries ALICE} Collaboration, K.~Aamodt {\em et~al.}, ``{Rapidity and
  transverse momentum dependence of inclusive J$/\psi$ production in pp
  collisions at $\sqrt{s} = 7$ TeV}'',
  \href{http://dx.doi.org/10.1016/j.physletb.2011.09.054,
  10.1016/j.physletb.2012.10.060}{{\em Phys. Lett.} {\bfseries B704} (2011)
  442--455}, \href{http://arxiv.org/abs/1105.0380}{{\ttfamily arXiv:1105.0380
  [hep-ex]}}.
[Erratum: Phys. Lett.B718,692(2012)].

\bibitem{Alice:2016wka}
{\bfseries ALICE} Collaboration, J.~Adam {\em et~al.}, ``{W and Z boson
  production in p--Pb collisions at $\sqrt{s_{\rm NN}}$ = 5.02 TeV}'',
  \href{http://dx.doi.org/10.1007/JHEP02(2017)077}{{\em JHEP} {\bfseries 02}
  (2017) 077}, \href{http://arxiv.org/abs/1611.03002}{{\ttfamily
  arXiv:1611.03002 [nucl-ex]}}.

\bibitem{Acharya:2019mky}
{\bfseries ALICE} Collaboration, S.~Acharya {\em et~al.}, ``{Production of
  muons from heavy-flavour hadron decays in pp collisions at $ \sqrt{s} $ =
  5.02 TeV}'', \href{http://dx.doi.org/10.1007/JHEP09(2019)008}{{\em JHEP}
  {\bfseries 09} (2019) 008}, \href{http://arxiv.org/abs/1905.07207}{{\ttfamily
  arXiv:1905.07207 [nucl-ex]}}.

\bibitem{Cacciari:1998it}
M.~Cacciari, M.~Greco, and P.~Nason, ``{The $p_T$ spectrum in heavy flavor
  hadroproduction}'',
  \href{http://dx.doi.org/10.1088/1126-6708/1998/05/007}{{\em JHEP} {\bfseries
  05} (1998) 007},
\href{http://arxiv.org/abs/hep-ph/9803400}{{\ttfamily arXiv:hep-ph/9803400
  [hep-ph]}}.

\bibitem{Adam:2015ptt}
{\bfseries ALICE} Collaboration, J.~Adam {\em et~al.}, ``{Centrality dependence
  of the charged-particle multiplicity density at midrapidity in Pb--Pb
  collisions at $\sqrt{s_{\rm NN}}$ = 5.02 TeV}'',
  \href{http://dx.doi.org/10.1103/PhysRevLett.116.222302}{{\em Phys. Rev.
  Lett.} {\bfseries 116} no.~22, (2016) 222302},
  \href{http://arxiv.org/abs/1512.06104}{{\ttfamily arXiv:1512.06104
  [nucl-ex]}}.

\bibitem{Acharya:2018hhy}
{\bfseries ALICE} Collaboration, S.~Acharya {\em et~al.}, ``{Centrality and
  pseudorapidity dependence of the charged-particle multiplicity density in
  Xe--Xe collisions at $\sqrt{s_{\rm NN}}$ =5.44TeV}'',
  \href{http://dx.doi.org/10.1016/j.physletb.2018.12.048}{{\em Phys. Lett. B}
  {\bfseries 790} (2019) 35--48},
  \href{http://arxiv.org/abs/1805.04432}{{\ttfamily arXiv:1805.04432
  [nucl-ex]}}.

\bibitem{Acharya:2019yoi}
{\bfseries ALICE} Collaboration, S.~Acharya {\em et~al.}, ``{Production of
  charged pions, kaons and (anti-)protons in Pb-Pb and inelastic pp collisions
  at $\sqrt{s_{\rm{NN}}}$ = 5.02 TeV}'',
  \href{http://dx.doi.org/10.1103/PhysRevC.101.044907}{{\em Phys. Rev. C}
  {\bfseries 101} no.~4, (2020) 044907},
  \href{http://arxiv.org/abs/1910.07678}{{\ttfamily arXiv:1910.07678
  [nucl-ex]}}.

\bibitem{Acharya:2018qsh}
{\bfseries ALICE} Collaboration, S.~Acharya {\em et~al.}, ``{Transverse
  momentum spectra and nuclear modification factors of charged particles in pp,
  p-Pb and Pb-Pb collisions at the LHC}'',
  \href{http://dx.doi.org/10.1007/JHEP11(2018)013}{{\em JHEP} {\bfseries 11}
  (2018) 013},
\href{http://arxiv.org/abs/1802.09145}{{\ttfamily arXiv:1802.09145 [nucl-ex]}}.

\bibitem{Sjostrand:2014zea}
T.~Sj{\"o}strand, S.~Ask, J.~R. Christiansen, R.~Corke, N.~Desai, P.~Ilten,
  S.~Mrenna, S.~Prestel, C.~O. Rasmussen, and P.~Z. Skands, ``{An Introduction
  to PYTHIA 8.2}'', \href{http://dx.doi.org/10.1016/j.cpc.2015.01.024}{{\em
  Comput. Phys. Commun.} {\bfseries 191} (2015) 159},
\href{http://arxiv.org/abs/1410.3012}{{\ttfamily arXiv:1410.3012 [hep-ph]}}.

\bibitem{Hijing:ref}
M.~Gyulassy and X.-N. Wang, ``{HIJING 1.0: A Monte Carlo program for parton and
  particle production in high-energy hadronic and nuclear collisions}'',
  \href{http://dx.doi.org/10.1016/0010-4655(94)90057-4}{{\em Comput. Phys.
  Commun.} {\bfseries 83} (1994) 307},
\href{http://arxiv.org/abs/nucl-th/9502021}{{\ttfamily arXiv:nucl-th/9502021
  [nucl-th]}}.

\bibitem{Brun:1082634}
R.~Brun, F.~Bruyant, F.~Carminati, S.~Giani, M.~Maire, A.~McPherson,
  G.~Patrick, and L.~Urban, ``{GEANT Detector Description and Simulation
  Tool}.'' {\href{10.17181/CERN.MUHF.DMJ1}{CERN-W5013, CERN-W-5013, W5013,
  W-5013}}, 1994.

\bibitem{Adam:2016rdg}
{\bfseries ALICE} Collaboration, J.~Adam {\em et~al.}, ``{J/$\psi$ suppression
  at forward rapidity in Pb-Pb collisions at $\mathbf{\sqrt{s_{{\rm NN}}} =
  5.02}$ TeV}'', \href{http://dx.doi.org/10.1016/j.physletb.2016.12.064}{{\em
  Phys. Lett.} {\bfseries B766} (2017) 212},
\href{http://arxiv.org/abs/1606.08197}{{\ttfamily arXiv:1606.08197 [nucl-ex]}}.

\bibitem{Acharya:2019iur}
{\bfseries ALICE} Collaboration, S.~Acharya {\em et~al.}, ``{Studies of
  J/$\psi$ production at forward rapidity in Pb-Pb collisions at
  $\sqrt{s_{\rm{NN}}}$ = 5.02 TeV}'',
  \href{http://dx.doi.org/10.1007/JHEP02(2020)041}{{\em JHEP} {\bfseries 02}
  (2020) 041}, \href{http://arxiv.org/abs/1909.03158}{{\ttfamily
  arXiv:1909.03158 [nucl-ex]}}.

\bibitem{Acharya:2018upq}
{\bfseries ALICE} Collaboration, S.~Acharya {\em et~al.}, ``{Measurements of
  low-$p_{\rm T}$ electrons from semileptonic heavy-flavour hadron decays at
  mid-rapidity in pp and Pb--Pb collisions at $ \sqrt{s_{\mathrm{NN}}}=2.76 $
  TeV}'', \href{http://dx.doi.org/10.1007/JHEP10(2018)061}{{\em JHEP}
  {\bfseries 10} (2018) 061},
\href{http://arxiv.org/abs/1805.04379}{{\ttfamily arXiv:1805.04379 [nucl-ex]}}.

\bibitem{Adam:2016khe}
{\bfseries ALICE} Collaboration, J.~Adam {\em et~al.}, ``{Measurement of the
  production of high-$p_{\rm T}$ electrons from heavy-flavour hadron decays in
  Pb--Pb collisions at $\mathbf{\sqrt{\it s_{\rm{NN}}}}$ = 2.76 TeV}'',
  \href{http://dx.doi.org/10.1016/j.physletb.2017.05.060}{{\em Phys. Lett.}
  {\bfseries B771} (2017) 467},
\href{http://arxiv.org/abs/1609.07104}{{\ttfamily arXiv:1609.07104 [nucl-ex]}}.

\bibitem{BetheBloch}
H.~Bethe, ``{Zur Theorie des Durchgangs schneller Korpuskularstrahlen durch
  Materie}'', \href{http://dx.doi.org/10.1002/andp.19303970303}{{\em Annalen
  der Physik} {\bfseries 397} no.~3, (1930) 325}.
  \url{http://dx.doi.org/10.1002/andp.19303970303}.

\bibitem{PhysRevD.98.030001}
{\bfseries Particle Data Group} Collaboration, ``Review of particle physics'',
  \href{http://dx.doi.org/10.1103/PhysRevD.98.030001}{{\em Phys. Rev. D}
  {\bfseries 98} (2018) 030001}.
  \url{https://link.aps.org/doi/10.1103/PhysRevD.98.030001}.

\bibitem{Acharya:2019hao}
{\bfseries ALICE} Collaboration, S.~Acharya {\em et~al.}, ``{Measurement of
  electrons from heavy-flavour hadron decays as a function of multiplicity in
  p--Pb collisions at $\sqrt{s_{\rm NN}}$ = 5.02 TeV}'',
  \href{http://dx.doi.org/10.1007/JHEP02(2020)077}{{\em JHEP} {\bfseries 02}
  (2020) 077},
\href{http://arxiv.org/abs/1910.14399}{{\ttfamily arXiv:1910.14399 [nucl-ex]}}.

\bibitem{Abelev:2012xe}
{\bfseries ALICE} Collaboration, B.~Abelev {\em et~al.}, ``{Measurement of
  electrons from semileptonic heavy-flavour hadron decays in pp collisions at
  $\sqrt{s}$ = 7 TeV}'',
  \href{http://dx.doi.org/10.1103/PhysRevD.86.112007}{{\em Phys. Rev.}
  {\bfseries D86} (2012) 112007},
\href{http://arxiv.org/abs/1205.5423}{{\ttfamily arXiv:1205.5423 [hep-ex]}}.

\bibitem{Khandai:2011cf}
P.~K. Khandai, P.~Shukla, and V.~Singh, ``{Meson spectra and $m_T$ scaling in
  $p + p$, $d + $Au, and Au + Au collisions at $\sqrt{s_{NN}}=200$ GeV}'',
  \href{http://dx.doi.org/10.1103/PhysRevC.84.054904}{{\em Phys. Rev.}
  {\bfseries C84} (2011) 054904},
\href{http://arxiv.org/abs/1110.3929}{{\ttfamily arXiv:1110.3929 [hep-ph]}}.

\bibitem{Altenkamper:2017qot}
L.~Altenkämper, F.~Bock, C.~Loizides, and N.~Schmidt, ``{Applicability of
  transverse mass scaling in hadronic collisions at energies available at the
  CERN Large Hadron Collider}'',
  \href{http://dx.doi.org/10.1103/PhysRevC.96.064907}{{\em Phys. Rev.}
  {\bfseries C96} (2017) 064907},
\href{http://arxiv.org/abs/1710.01933}{{\ttfamily arXiv:1710.01933 [hep-ph]}}.

\bibitem{Adam:2015kca}
{\bfseries ALICE} Collaboration, J.~Adam {\em et~al.}, ``{Centrality dependence
  of the nuclear modification factor of charged pions, kaons, and protons in
  Pb--Pb collisions at $\sqrt{s_{\rm NN}}=2.76$ TeV}'',
  \href{http://dx.doi.org/10.1103/PhysRevC.93.034913}{{\em Phys. Rev. C}
  {\bfseries 93} (2016) 034913},
  \href{http://arxiv.org/abs/1506.07287}{{\ttfamily arXiv:1506.07287
  [nucl-ex]}}.

\bibitem{Sjostrand:2006za}
T.~Sj{\"o}strand, S.~Mrenna, and P.~Z. Skands, ``{PYTHIA 6.4 Physics and
  Manual}'', \href{http://dx.doi.org/10.1088/1126-6708/2006/05/026}{{\em JHEP}
  {\bfseries 05} (2006) 026},
\href{http://arxiv.org/abs/hep-ph/0603175}{{\ttfamily arXiv:hep-ph/0603175
  [hep-ph]}}.

\bibitem{Abelev:2012sca}
{\bfseries ALICE} Collaboration, B.~Abelev {\em et~al.}, ``{Measurement of
  electrons from beauty hadron decays in pp collisions at $\sqrt{s}=7$ TeV}'',
  \href{http://dx.doi.org/10.1016/j.physletb.2013.01.069}{{\em Phys. Lett.}
  {\bfseries B721} (2013) 13},
\href{http://arxiv.org/abs/1208.1902}{{\ttfamily arXiv:1208.1902 [hep-ex]}}.

\bibitem{Christiansen:2015yqa}
J.~R. Christiansen and P.~Z. Skands, ``{String Formation Beyond Leading
  Colour}'', \href{http://dx.doi.org/10.1007/JHEP08(2015)003}{{\em JHEP}
  {\bfseries 08} (2015) 003}, \href{http://arxiv.org/abs/1505.01681}{{\ttfamily
  arXiv:1505.01681 [hep-ph]}}.

\bibitem{Averbeck:2011ga}
R.~Averbeck, N.~Bastid, Z.~C. del Valle, P.~Crochet, A.~Dainese, and X.~Zhang,
  ``{Reference Heavy Flavour Cross Sections in pp Collisions at $\sqrt{s}$ =
  2.76 TeV, using a pQCD-Driven $\sqrt{s}$-Scaling of ALICE Measurements at
  $\sqrt s$ s = 7 TeV}'',
\href{http://arxiv.org/abs/1107.3243}{{\ttfamily arXiv:1107.3243 [hep-ph]}}.

\bibitem{ALICE-PUBLIC-2018-014}
{\bfseries ALICE Collaboration} Collaboration, ``{ALICE 2017 luminosity
  determination for pp collisions at $\sqrt{s}$ = 5 TeV}.''
  {ALICE-PUBLIC-2018-014}, 2018.
\newblock \url{http://cds.cern.ch/record/2648933}.

\bibitem{ALICE-PUBLIC-2016-005}
{\bfseries ALICE Collaboration} Collaboration, ``{ALICE luminosity
  determination for pp collisions at $\sqrt{s}=5$ TeV}.''
  {ALICE-PUBLIC-2018-011}, 2016.
\newblock \url{https://cds.cern.ch/record/2202638}.

\bibitem{ALICE-PUBLIC-2018-005}
{\bfseries ALICE Collaboration} Collaboration, ``{Preliminary Physics Summary:
  Measurements of low-$p_{\rm T}$ electrons from semileptonic heavy-flavour
  hadron decays at mid-rapidity in pp collisions at $\sqrt{s}= 7$ TeV}.''
  {ALICE-PUBLIC-2018-005}, 2018.
\newblock \url{https://cds.cern.ch/record/2317185}.

\bibitem{Song:2015ykw}
T.~Song, H.~Berrehrah, D.~Cabrera, W.~Cassing, and E.~Bratkovskaya, ``{Charm
  production in Pb+Pb collisions at energies available at the CERN Large Hadron
  Collider}'', \href{http://dx.doi.org/10.1103/PhysRevC.93.034906}{{\em Phys.
  Rev.} {\bfseries C93} no.~3, (2016) 034906},
\href{http://arxiv.org/abs/1512.00891}{{\ttfamily arXiv:1512.00891 [nucl-th]}}.

\bibitem{Nahrgang:2013xaa}
M.~Nahrgang, J.~Aichelin, P.~B. Gossiaux, and K.~Werner, ``{Influence of
  hadronic bound states above $T_{\rm c}$ on heavy-quark observables in
  {{Pb--Pb}} collisions at at the CERN Large Hadron Collider}'',
  \href{http://dx.doi.org/10.1103/PhysRevC.89.014905}{{\em Phys.Rev.}
  {\bfseries C89} (2014) 014905},
\href{http://arxiv.org/abs/1305.6544}{{\ttfamily arXiv:1305.6544 [hep-ph]}}.

\bibitem{Djordjevic:2015hra}
M.~Djordjevic and M.~Djordjevic, ``{Predictions of heavy-flavor suppression at
  5.1 TeV Pb+Pb collisions at the CERN Large Hadron Collider}'',
  \href{http://dx.doi.org/10.1103/PhysRevC.92.024918}{{\em Phys. Rev.}
  {\bfseries C92} (2015) 024918},
\href{http://arxiv.org/abs/1505.04316}{{\ttfamily arXiv:1505.04316 [nucl-th]}}.

\bibitem{Djordjevic:2014}
M.~Djordjevic and M.~Djordjevic, ``{LHC jet suppression of light and heavy
  flavor observables}'',
  \href{http://dx.doi.org/10.1016/j.physletb.2014.05.053}{{\em Phys. Lett.}
  {\bfseries B734} (2014) 286},
\href{http://arxiv.org/abs/1307.4098}{{\ttfamily arXiv:1307.4098 [hep-ph]}}.

\bibitem{Acharya:2017hdv}
{\bfseries ALICE} Collaboration, S.~Acharya {\em et~al.}, ``{Production of
  muons from heavy-flavour hadron decays in p--Pb collisions at
  $\mathbf{\sqrt{{\textit s}_{NN}} = 5.02}$ TeV}'',
  \href{http://dx.doi.org/10.1016/j.physletb.2017.03.049}{{\em Phys. Lett.}
  {\bfseries B770} (2017) 459},
\href{http://arxiv.org/abs/1702.01479}{{\ttfamily arXiv:1702.01479 [nucl-ex]}}.

\bibitem{Adam:2015sza}
{\bfseries ALICE} Collaboration, J.~Adam {\em et~al.}, ``{Transverse momentum
  dependence of D-meson production in Pb-Pb collisions at $
  \sqrt{{\mathrm{s}}_{\mathrm{NN}}}=$ 2.76 TeV}'',
  \href{http://dx.doi.org/10.1007/JHEP03(2016)081}{{\em JHEP} {\bfseries 03}
  (2016) 081},
\href{http://arxiv.org/abs/1509.06888}{{\ttfamily arXiv:1509.06888 [nucl-ex]}}.

\bibitem{PhysRevD.11.3105}
J.~W. Cronin, H.~J. Frisch, M.~J. Shochet, J.~P. Boymond, P.~A. Pirou\'e, and
  R.~L. Sumner, ``Production of hadrons at large transverse momentum at 200,
  300, and 400 gev'', \href{http://dx.doi.org/10.1103/PhysRevD.11.3105}{{\em
  Phys. Rev.} {\bfseries D11} (1975) 3105}.
  \url{https://link.aps.org/doi/10.1103/PhysRevD.11.3105}.

\bibitem{dEnterria:2009xfs}
D.~d'Enterria, {\em {Jet quenching}}, vol.~23,
  \href{http://dx.doi.org/10.1007/978-3-642-01539-7\_16}{p.~471}.
\newblock 2010.
\newblock \href{http://arxiv.org/abs/0902.2011}{{\ttfamily arXiv:0902.2011
  [nucl-ex]}}.

\bibitem{Chatrchyan:2012mb}
{\bfseries CMS} Collaboration, S.~Chatrchyan {\em et~al.}, ``{Measurement of
  the pseudorapidity and centrality dependence of the transverse energy density
  in PbPb collisions at $\sqrt{s_{NN}}=2.76$ TeV}'',
  \href{http://dx.doi.org/10.1103/PhysRevLett.109.152303}{{\em Phys. Rev.
  Lett.} {\bfseries 109} (2012) 152303},
  \href{http://arxiv.org/abs/1205.2488}{{\ttfamily arXiv:1205.2488 [nucl-ex]}}.

\bibitem{Baier:1996kr}
R.~Baier, Y.~L. Dokshitzer, A.~H. Mueller, S.~Peigne, and D.~Schiff,
  ``{Radiative energy loss of high-energy quarks and gluons in a finite volume
  quark - gluon plasma}'',
  \href{http://dx.doi.org/10.1016/S0550-3213(96)00553-6}{{\em Nucl.Phys.}
  {\bfseries B483} (1997) 291},
\href{http://arxiv.org/abs/hep-ph/9607355}{{\ttfamily arXiv:hep-ph/9607355
  [hep-ph]}}.

\bibitem{ALICE-PUBLIC-2020-009}
{\bfseries ALICE} Collaboration, ``{Supplemental figure: Production of muons
  from heavy-flavour hadron decays at high transverse momentum in Pb-Pb
  collisions at $\sqrt{s_{\rm NN}} = 5.02$ and $2.76$ TeV}'', 2020.
\newblock \url{https://cds.cern.ch/record/2747060}.

\bibitem{Adare:2015cua}
{\bfseries PHENIX} Collaboration, A.~Adare {\em et~al.}, ``{Scaling properties
  of fractional momentum loss of high-$p_T$ hadrons in nucleus-nucleus
  collisions at $\sqrt{s_{_{NN}}}$ from 62.4 GeV to 2.76 TeV}'',
  \href{http://dx.doi.org/10.1103/PhysRevC.93.024911}{{\em Phys. Rev. C}
  {\bfseries 93} no.~2, (2016) 024911},
  \href{http://arxiv.org/abs/1509.06735}{{\ttfamily arXiv:1509.06735
  [nucl-ex]}}.

\end{thebibliography}\endgroup

\newpage
\appendix

\section{The ALICE Collaboration}
\label{app:collab}
\begingroup
\small
\begin{flushleft}


S.~Acharya$^{\rm 142}$, 
D.~Adamov\'{a}$^{\rm 97}$, 
A.~Adler$^{\rm 75}$, 
J.~Adolfsson$^{\rm 82}$, 
G.~Aglieri Rinella$^{\rm 35}$, 
M.~Agnello$^{\rm 31}$, 
N.~Agrawal$^{\rm 55}$, 
Z.~Ahammed$^{\rm 142}$, 
S.~Ahmad$^{\rm 16}$, 
S.U.~Ahn$^{\rm 77}$
% ...

\section*{Affiliation notes}

$^{\rm I}$ Deceased\\
$^{\rm II}$ Also at: Italian National Agency for New Technologies, Energy and Sustainable Economic Development (ENEA), Bologna, Italy\\
$^{\rm III}$ Also at: Dipartimento DET del Politecnico di Torino, Turin, Italy\\
$^{\rm IV}$ Also at: M.V. Lomonosov Moscow State University, D.V. Skobeltsyn Institute of Nuclear, Physics, Moscow, Russia\\
$^{\rm V}$ Also at: Institute of Theoretical Physics, University of Wroclaw, Poland\\

\section*{Collaboration Institutes}

$^{1}$ A.I. Alikhanyan National Science Laboratory (Yerevan Physics Institute) Foundation, Yerevan, Armenia\\
$^{2}$ AGH University of Science and Technology, Cracow, Poland\\
$^{3}$ Bogolyubov Institute for Theoretical Physics, National Academy of Sciences of Ukraine, Kiev, Ukraine\\
$^{4}$ Bose Institute, Department of Physics  and Centre for Astroparticle Physics and Space Science (CAPSS), Kolkata, India\\
$^{5}$ Budker Institute for Nuclear Physics, Novosibirsk, Russia\\
$^{6}$ California Polytechnic State University, San Luis Obispo, California, United States\\
$^{7}$ Central China Normal University, Wuhan, China\\
$^{8}$ Centro de Aplicaciones Tecnol\'{o}gicas y Desarrollo Nuclear (CEADEN), Havana, Cuba\\
$^{9}$ Centro de Investigaci\'{o}n y de Estudios Avanzados (CINVESTAV), Mexico City and M\'{e}rida, Mexico\\
$^{10}$ Chicago State University, Chicago, Illinois, United States\\
$^{11}$ China Institute of Atomic Energy, Beijing, China\\
$^{12}$ Chungbuk National University, Cheongju, Republic of Korea\\
$^{13}$ Comenius University Bratislava, Faculty of Mathematics, Physics and Informatics, Bratislava, Slovakia\\
$^{14}$ COMSATS University Islamabad, Islamabad, Pakistan\\
$^{15}$ Creighton University, Omaha, Nebraska, United States\\
$^{16}$ Department of Physics, Aligarh Muslim University, Aligarh, India\\
$^{17}$ Department of Physics, Pusan National University, Pusan, Republic of Korea\\
$^{18}$ Department of Physics, Sejong University, Seoul, Republic of Korea\\
$^{19}$ Department of Physics, University of California, Berkeley, California, United States\\
$^{20}$ Department of Physics, University of Oslo, Oslo, Norway\\
$^{21}$ Department of Physics and Technology, University of Bergen, Bergen, Norway\\
$^{22}$ Dipartimento di Fisica dell'Universit\`{a} 'La Sapienza' and Sezione INFN, Rome, Italy\\
$^{23}$ Dipartimento di Fisica dell'Universit\`{a} and Sezione INFN, Cagliari, Italy\\
$^{24}$ Dipartimento di Fisica dell'Universit\`{a} and Sezione INFN, Trieste, Italy\\
$^{25}$ Dipartimento di Fisica dell'Universit\`{a} and Sezione INFN, Turin, Italy\\
$^{26}$ Dipartimento di Fisica e Astronomia dell'Universit\`{a} and Sezione INFN, Bologna, Italy\\
$^{27}$ Dipartimento di Fisica e Astronomia dell'Universit\`{a} and Sezione INFN, Catania, Italy\\
$^{28}$ Dipartimento di Fisica e Astronomia dell'Universit\`{a} and Sezione INFN, Padova, Italy\\
$^{29}$ Dipartimento di Fisica e Nucleare e Teorica, Universit\`{a} di Pavia  and Sezione INFN, Pavia, Italy\\
$^{30}$ Dipartimento di Fisica `E.R.~Caianiello' dell'Universit\`{a} and Gruppo Collegato INFN, Salerno, Italy\\
$^{31}$ Dipartimento DISAT del Politecnico and Sezione INFN, Turin, Italy\\
$^{32}$ Dipartimento di Scienze e Innovazione Tecnologica dell'Universit\`{a} del Piemonte Orientale and INFN Sezione di Torino, Alessandria, Italy\\
$^{33}$ Dipartimento di Scienze MIFT, Universit\`{a} di Messina, Messina, Italy\\
$^{34}$ Dipartimento Interateneo di Fisica `M.~Merlin' and Sezione INFN, Bari, Italy\\
$^{35}$ European Organization for Nuclear Research (CERN), Geneva, Switzerland\\
$^{36}$ Faculty of Electrical Engineering, Mechanical Engineering and Naval Architecture, University of Split, Split, Croatia\\
$^{37}$ Faculty of Engineering and Science, Western Norway University of Applied Sciences, Bergen, Norway\\
$^{38}$ Faculty of Nuclear Sciences and Physical Engineering, Czech Technical University in Prague, Prague, Czech Republic\\
$^{39}$ Faculty of Science, P.J.~\v{S}af\'{a}rik University, Ko\v{s}ice, Slovakia\\
$^{40}$ Frankfurt Institute for Advanced Studies, Johann Wolfgang Goethe-Universit\"{a}t Frankfurt, Frankfurt, Germany\\
$^{41}$ Fudan University, Shanghai, China\\
$^{42}$ Gangneung-Wonju National University, Gangneung, Republic of Korea\\
$^{43}$ Gauhati University, Department of Physics, Guwahati, India\\
$^{44}$ Helmholtz-Institut f\"{u}r Strahlen- und Kernphysik, Rheinische Friedrich-Wilhelms-Universit\"{a}t Bonn, Bonn, Germany\\
$^{45}$ Helsinki Institute of Physics (HIP), Helsinki, Finland\\
$^{46}$ High Energy Physics Group,  Universidad Aut\'{o}noma de Puebla, Puebla, Mexico\\
$^{47}$ Hiroshima University, Hiroshima, Japan\\
$^{48}$ Hochschule Worms, Zentrum  f\"{u}r Technologietransfer und Telekommunikation (ZTT), Worms, Germany\\
$^{49}$ Horia Hulubei National Institute of Physics and Nuclear Engineering, Bucharest, Romania\\
$^{50}$ Indian Institute of Technology Bombay (IIT), Mumbai, India\\
$^{51}$ Indian Institute of Technology Indore, Indore, India\\
$^{52}$ Indonesian Institute of Sciences, Jakarta, Indonesia\\
$^{53}$ INFN, Laboratori Nazionali di Frascati, Frascati, Italy\\
$^{54}$ INFN, Sezione di Bari, Bari, Italy\\
$^{55}$ INFN, Sezione di Bologna, Bologna, Italy\\
$^{56}$ INFN, Sezione di Cagliari, Cagliari, Italy\\
$^{57}$ INFN, Sezione di Catania, Catania, Italy\\
$^{58}$ INFN, Sezione di Padova, Padova, Italy\\
$^{59}$ INFN, Sezione di Roma, Rome, Italy\\
$^{60}$ INFN, Sezione di Torino, Turin, Italy\\
$^{61}$ INFN, Sezione di Trieste, Trieste, Italy\\
$^{62}$ Inha University, Incheon, Republic of Korea\\
$^{63}$ Institute for Gravitational and Subatomic Physics (GRASP), Utrecht University/Nikhef, Utrecht, Netherlands\\
$^{64}$ Institute for Nuclear Research, Academy of Sciences, Moscow, Russia\\
$^{65}$ Institute of Experimental Physics, Slovak Academy of Sciences, Ko\v{s}ice, Slovakia\\
$^{66}$ Institute of Physics, Homi Bhabha National Institute, Bhubaneswar, India\\
$^{67}$ Institute of Physics of the Czech Academy of Sciences, Prague, Czech Republic\\
$^{68}$ Institute of Space Science (ISS), Bucharest, Romania\\
$^{69}$ Institut f\"{u}r Kernphysik, Johann Wolfgang Goethe-Universit\"{a}t Frankfurt, Frankfurt, Germany\\
$^{70}$ Instituto de Ciencias Nucleares, Universidad Nacional Aut\'{o}noma de M\'{e}xico, Mexico City, Mexico\\
$^{71}$ Instituto de F\'{i}sica, Universidade Federal do Rio Grande do Sul (UFRGS), Porto Alegre, Brazil\\
$^{72}$ Instituto de F\'{\i}sica, Universidad Nacional Aut\'{o}noma de M\'{e}xico, Mexico City, Mexico\\
$^{73}$ iThemba LABS, National Research Foundation, Somerset West, South Africa\\
$^{74}$ Jeonbuk National University, Jeonju, Republic of Korea\\
$^{75}$ Johann-Wolfgang-Goethe Universit\"{a}t Frankfurt Institut f\"{u}r Informatik, Fachbereich Informatik und Mathematik, Frankfurt, Germany\\
$^{76}$ Joint Institute for Nuclear Research (JINR), Dubna, Russia\\
$^{77}$ Korea Institute of Science and Technology Information, Daejeon, Republic of Korea\\
$^{78}$ KTO Karatay University, Konya, Turkey\\
$^{79}$ Laboratoire de Physique des 2 Infinis, Ir\`{e}ne Joliot-Curie, Orsay, France\\
$^{80}$ Laboratoire de Physique Subatomique et de Cosmologie, Universit\'{e} Grenoble-Alpes, CNRS-IN2P3, Grenoble, France\\
$^{81}$ Lawrence Berkeley National Laboratory, Berkeley, California, United States\\
$^{82}$ Lund University Department of Physics, Division of Particle Physics, Lund, Sweden\\
$^{83}$ Moscow Institute for Physics and Technology, Moscow, Russia\\
$^{84}$ Nagasaki Institute of Applied Science, Nagasaki, Japan\\
$^{85}$ Nara Women{'}s University (NWU), Nara, Japan\\
$^{86}$ National and Kapodistrian University of Athens, School of Science, Department of Physics , Athens, Greece\\
$^{87}$ National Centre for Nuclear Research, Warsaw, Poland\\
$^{88}$ National Institute of Science Education and Research, Homi Bhabha National Institute, Jatni, India\\
$^{89}$ National Nuclear Research Center, Baku, Azerbaijan\\
$^{90}$ National Research Centre Kurchatov Institute, Moscow, Russia\\
$^{91}$ Niels Bohr Institute, University of Copenhagen, Copenhagen, Denmark\\
$^{92}$ Nikhef, National institute for subatomic physics, Amsterdam, Netherlands\\
$^{93}$ NRC Kurchatov Institute IHEP, Protvino, Russia\\
$^{94}$ NRC \guillemotleft Kurchatov\guillemotright  Institute - ITEP, Moscow, Russia\\
$^{95}$ NRNU Moscow Engineering Physics Institute, Moscow, Russia\\
$^{96}$ Nuclear Physics Group, STFC Daresbury Laboratory, Daresbury, United Kingdom\\
$^{97}$ Nuclear Physics Institute of the Czech Academy of Sciences, \v{R}e\v{z} u Prahy, Czech Republic\\
$^{98}$ Oak Ridge National Laboratory, Oak Ridge, Tennessee, United States\\
$^{99}$ Ohio State University, Columbus, Ohio, United States\\
$^{100}$ Petersburg Nuclear Physics Institute, Gatchina, Russia\\
$^{101}$ Physics department, Faculty of science, University of Zagreb, Zagreb, Croatia\\
$^{102}$ Physics Department, Panjab University, Chandigarh, India\\
$^{103}$ Physics Department, University of Jammu, Jammu, India\\
$^{104}$ Physics Department, University of Rajasthan, Jaipur, India\\
$^{105}$ Physikalisches Institut, Eberhard-Karls-Universit\"{a}t T\"{u}bingen, T\"{u}bingen, Germany\\
$^{106}$ Physikalisches Institut, Ruprecht-Karls-Universit\"{a}t Heidelberg, Heidelberg, Germany\\
$^{107}$ Physik Department, Technische Universit\"{a}t M\"{u}nchen, Munich, Germany\\
$^{108}$ Politecnico di Bari and Sezione INFN, Bari, Italy\\
$^{109}$ Research Division and ExtreMe Matter Institute EMMI, GSI Helmholtzzentrum f\"ur Schwerionenforschung GmbH, Darmstadt, Germany\\
$^{110}$ Rudjer Bo\v{s}kovi\'{c} Institute, Zagreb, Croatia\\
$^{111}$ Russian Federal Nuclear Center (VNIIEF), Sarov, Russia\\
$^{112}$ Saha Institute of Nuclear Physics, Homi Bhabha National Institute, Kolkata, India\\
$^{113}$ School of Physics and Astronomy, University of Birmingham, Birmingham, United Kingdom\\
$^{114}$ Secci\'{o}n F\'{\i}sica, Departamento de Ciencias, Pontificia Universidad Cat\'{o}lica del Per\'{u}, Lima, Peru\\
$^{115}$ St. Petersburg State University, St. Petersburg, Russia\\
$^{116}$ Stefan Meyer Institut f\"{u}r Subatomare Physik (SMI), Vienna, Austria\\
$^{117}$ SUBATECH, IMT Atlantique, Universit\'{e} de Nantes, CNRS-IN2P3, Nantes, France\\
$^{118}$ Suranaree University of Technology, Nakhon Ratchasima, Thailand\\
$^{119}$ Technical University of Ko\v{s}ice, Ko\v{s}ice, Slovakia\\
$^{120}$ The Henryk Niewodniczanski Institute of Nuclear Physics, Polish Academy of Sciences, Cracow, Poland\\
$^{121}$ The University of Texas at Austin, Austin, Texas, United States\\
$^{122}$ Universidad Aut\'{o}noma de Sinaloa, Culiac\'{a}n, Mexico\\
$^{123}$ Universidade de S\~{a}o Paulo (USP), S\~{a}o Paulo, Brazil\\
$^{124}$ Universidade Estadual de Campinas (UNICAMP), Campinas, Brazil\\
$^{125}$ Universidade Federal do ABC, Santo Andre, Brazil\\
$^{126}$ University of Cape Town, Cape Town, South Africa\\
$^{127}$ University of Houston, Houston, Texas, United States\\
$^{128}$ University of Jyv\"{a}skyl\"{a}, Jyv\"{a}skyl\"{a}, Finland\\
$^{129}$ University of Liverpool, Liverpool, United Kingdom\\
$^{130}$ University of Science and Technology of China, Hefei, China\\
$^{131}$ University of South-Eastern Norway, Tonsberg, Norway\\
$^{132}$ University of Tennessee, Knoxville, Tennessee, United States\\
$^{133}$ University of the Witwatersrand, Johannesburg, South Africa\\
$^{134}$ University of Tokyo, Tokyo, Japan\\
$^{135}$ University of Tsukuba, Tsukuba, Japan\\
$^{136}$ Universit\'{e} Clermont Auvergne, CNRS/IN2P3, LPC, Clermont-Ferrand, France\\
$^{137}$ Universit\'{e} de Lyon, CNRS/IN2P3, Institut de Physique des 2 Infinis de Lyon , Lyon, France\\
$^{138}$ Universit\'{e} de Strasbourg, CNRS, IPHC UMR 7178, F-67000 Strasbourg, France, Strasbourg, France\\
$^{139}$ Universit\'{e} Paris-Saclay Centre d'Etudes de Saclay (CEA), IRFU, D\'{e}partment de Physique Nucl\'{e}aire (DPhN), Saclay, France\\
$^{140}$ Universit\`{a} degli Studi di Foggia, Foggia, Italy\\
$^{141}$ Universit\`{a} di Brescia and Sezione INFN, Brescia, Italy\\
$^{142}$ Variable Energy Cyclotron Centre, Homi Bhabha National Institute, Kolkata, India\\
$^{143}$ Warsaw University of Technology, Warsaw, Poland\\
$^{144}$ Wayne State University, Detroit, Michigan, United States\\
$^{145}$ Westf\"{a}lische Wilhelms-Universit\"{a}t M\"{u}nster, Institut f\"{u}r Kernphysik, M\"{u}nster, Germany\\
$^{146}$ Wigner Research Centre for Physics, Budapest, Hungary\\
$^{147}$ Yale University, New Haven, Connecticut, United States\\
$^{148}$ Yonsei University, Seoul, Republic of Korea\\

\endgroup 
\end{document}